# The Measurement Problem: Decoherence and Convivial Solipsism[1]

(Long version)

HERVE ZWIRN

*UFR de Physique (LIED, Université Paris 7), CMLA (ENS Cachan, France) & IHPST (CNRS, France)*

*herve.zwirn@gmail.com*

## ABSTRACT

The problem of measurement is often considered as an inconsistency inside the quantum formalism. Many attempts to solve (or to dissolve) it have been made since the inception of quantum mechanics. The form of these attempts depends on the philosophical position that their authors endorse. I will review some of them and analyze their relevance. The phenomenon of decoherence is often presented as a solution lying inside the pure quantum formalism and not demanding any particular philosophical assumption. Nevertheless, a widely debated question is to decide between two different interpretations. The first one is to consider that the decoherence process has the effect to actually project a superposed state into one of its classically interpretable component, hence doing the same job as the reduction postulate. For the second one, decoherence is only a way to show why no macroscopic superposed state can be observed, so explaining the classical appearance of the macroscopic world, while the quantum entanglement between the system, the apparatus and the environment never disappears. In this case, explaining why only one single definite outcome is observed remains to do. In this paper, I examine the arguments that have been given for and against both interpretations and defend a new position, the "Convivial Solipsism", according to which the outcome that is observed is relative to the observer, different but in close parallel to the Everett's interpretation and sharing also some similarities with Rovelli's relational interpretation and Quantum Bayesianism. I also show how "Convivial Solipsism" can help getting a new standpoint about the EPR paradox providing a way out of the standard dilemma that is having to choose between abandoning either realism or locality.

## 1. Introduction

Quantum mechanics works extremely well to predict the results of measurement. No example of conflict between experiments and predictions has been found yet. So, if you are interested in predictions of experiments and only in predictions of experiments, you can stop here. But if you are wondering just a little bit more about the formalism that is used to make these predictions then you open the door to a series of questions that seem not easy at all to answer. For example, you could be interested in knowing whether quantum formalism is only a list of recipes, the terms it contains having no other meaning than the one directly linked to the results of experiments, or if these terms mean something more, being related in some way to the real world that we seem to live in. You could think that the formalism is nothing else but the simple reflection of your knowledge of the system. You could as well be interested in knowing whether it is possible to give a description of what happens "between" experiments or if this question is meaningless. Trying to answer to only one of these questions leads to examine a whole set of analogous questions that are intertwined into a complex network of relationship. One central question that is at the core of all the others is the measurement problem.

---

[1] This paper is a widely extended version of a conference I gave at the 14th annual international symposium "Frontiers of Fundamental Physics" (FFP14) in Marseille (France) in July 2014 [1].

Although I don't share his realist point of view when Bell says [2]:

*"But it is interesting to speculate on the possibility that a future theory will not be intrinsically ambiguous and approximate. Such a theory could not be fundamentally about measurements for that would again imply incompleteness of the system and unanalyzed interventions from the outside. Rather it should again become possible to say of a system not that such and such may be observed but rather that such and such be so."*

and that I think, on the contrary, that it is unavoidable that the concept of measurement be deeply rooted inside any quantum theory as its main object, I think that Bell is one of those who have described in the most accurate and explicit way the lacuna of the commonly agreed presentations of the quantum formalism. It is an illusion to believe that the process of measurement is correctly described in the usual treatments such as provided in the majority of textbooks (except if it is only FAPP i.e. "for all practical purposes") and Bell shows that very clearly [3]:

*"The first charge against 'measurement', in the fundamental axioms of quantum mechanics, is that it anchors there the shifty split of the world into 'system' and 'apparatus'. [….] It remains that the theory is ambiguous in principle, about exactly when and exactly how the collapse occurs, about what is microscopic and what is macroscopic, what quantum and what classical."*

and:

*"What exactly qualifies some physical systems to play the role of 'measurer'? Was the wave function of the world waiting to jump for thousands of millions of year until a single-celled living creature appeared? Or did it have to wait a little longer, for some better qualified system … with a PhD?"*

The measurement problem lies at the very heart of the conceptual problems of interpretation of quantum mechanics. Depending on the stance they adopt, physicists put up with the difficulty more or less easily. The problem is more acute for those who endorse a realist position assuming that the wave function (or the state vector) represents the real state of the system than for those, as instrumentalists, who think that the formalism is but a tool allowing us to make predictions about future results of experiments. For physicists like John Bell, who want to maintain the view that physics must describe things as they really are, the problem is so serious that he says [3] he can't be satisfied with standard quantum mechanics (even if he acknowledges that it is working perfectly well for all practical purposes). In this case, quantum mechanics is not considered as complete and must be either completed by hidden variables or modified to welcome a dynamical process allowing for the state reduction. Another possibility is also to admit that the quantum formalism is not universal and that there are "entities" for which it is not applicable. On the contrary, for Bohr and the supporters of the so-called Copenhagen interpretation, quantum mechanics is complete and there is no problem if a correct epistemological distinction between the apparatus and the system is made[2]. Nevertheless, whatever position is adopted, it is not so easy to completely get rid of the problem despite what some authors seem to have claimed.

In this paper, I first state the problem under its most usual form and then I examine what, in this formulation, is directly related to hidden assumptions that could be eliminated in order to avoid the difficulties. I then show that many previously proposed solutions are not entirely satisfying and replace them in the philosophical context on which they rest. I present the consequences of the EPR paradox, Kochen Specher theorem and Bell's inequalities that must be taken into account if one wants to provide a coherent and acceptable interpretation. I give a series of puzzling questions to which any serious interpretation should reply and I stress the fact that it is the case for no previous interpretation. I analyze then the decoherence program which has often been understood as providing a solution of the measurement problem using only standard rules of the quantum formalism. I show that this is not the case. I conclude that, if we want to stay inside the standard quantum formalism (without switching to hidden variables theories or involving some modification of the Schrödinger equation) it seems impossible to define what a measurement is without any reference to a conscious observer. I give then a presentation of the so-called "Convivial Solipsism" which aims at switching to a point of view according to which it is unavoidable that the observer be placed at the center of the process of measurement. I show how Convivial Solipsism can help understanding the EPR paradox in a new way, providing a path out of the standard dilemma of having to choose between abandoning either realism or locality. At last, I argue to say that Convivial Solipsism is giving a global coherent interpretation answering every question of the series of puzzles.

---

[2] Of course this is an oversimplified presentation of Bohr's position which will be discussed in §3.2 in much more details.

In order for this paper to be self-sufficient and at the price of being a little bit longer than it could be possible would I assume the reader to be already familiar with all these results, I will present explicitly all is needed to understand which questions beg for an answer and why Convivial Solipsism provides an interpretation answering them.

## 2. The measurement problem

### 2.1. Formulation of the problem

Quantum formalism contains two postulates for computing the evolution of the state of a system. The first one is the Schrödinger equation: $i\hbar \frac{d|\Psi\rangle}{dt} = H|\Psi\rangle$ which is supposed to be used for an isolated system when no measurement is performed on it. The second one is the reduction postulate which says that when a measurement of a certain observable P is made on a system which is initially in a state that is a superposition of eingenstates of P, $|\Psi\rangle = \sum c_i |\varphi_i\rangle$, then after the measurement, if the result is $\lambda_k$, one eigenvalue of P, the state $|\Psi\rangle$ is projected onto the eigenvector $|\varphi_k\rangle$ linked to this eigenvalue $\sum c_i |\varphi_i\rangle \rightarrow |\varphi_k\rangle$ or onto the sub space of the Hilbert space that is spanned by the eigenvectors linked to $\lambda_k$, if $\lambda_k$ is degenerated.

Now these two computations do not lead to the same result. We follow here the notorious analysis given by von Neumann [4]. A measurement is an interaction between a system and an apparatus. Let the system S be in a state $|\Psi_S\rangle = \sum c_i |\varphi_i\rangle$ and the apparatus A be in the initial state $|A_0\rangle$. Then, before they interact, the state of the system – apparatus is the tensorial product

$$|\Psi_{SA}\rangle = |\Psi_S\rangle \otimes |A_0\rangle = \sum c_i |\varphi_i\rangle \otimes |A_0\rangle \qquad (1)$$

The interaction between S and A is done through a Hamiltonian $H^{AS}$ operating during a short time. It is assumed that the apparatus is built in such a way that if the measurement is made on a system that is in the state $|\varphi_i\rangle$, the apparatus will be in the state $|A_i\rangle$ after the measurement whatever its initial state.

Let's consider that a measurement is made on the system S and that a value $\lambda_k$ is found (let's suppose that it is not degenerated). The reduction postulate gives:

$$|\Psi_S\rangle = \sum c_i |\varphi_i\rangle \rightarrow |\varphi_k\rangle \text{ and } |A_0\rangle \rightarrow |A_k\rangle \qquad (2)$$

This is the usual description of a measurement. The apparatus is observed in the state $|A_k\rangle$ which is correlated to $|\varphi_k\rangle$ and it is hence inferred that the value of P for the system is $\lambda_k$ and that the state of the system is projected onto the state $|\varphi_k\rangle$. For example, if the system is a spin ½ particle and the apparatus a detector with a needle such as a spin +1/2 along Oz leads to a position of the needle pointing up (and a spin -1/2, a position of the needle pointing down) then equation (2) means that at the end of the measurement the needle is either pointing up and the spin of the particle is +1/2 or pointing down and the spin of the particle is -1/2 whatever the initial state of the particle be.

But it is as well possible to consider the system-apparatus as an isolated global system on which no measurement is made. In this case, the Schrodinger equation which describes a linear and unitary process gives:

$$|\Psi_{SA}\rangle = \sum c_i |\varphi_i\rangle \otimes |A_0\rangle \to \sum c_i |\varphi_i\rangle \otimes |A_i\rangle \quad (3)$$

Equation (3) shows that after the interaction, the system and the apparatus are in an entangled state. In particular, this is to be interpreted as if the apparatus was in a state which is a superposition of states linked to different possible results of the measurement[3]. In the above example of a spin ½ particle, equation (3) leads to a state of the needle that is a superposition of positions up and down. Of course, no such macroscopic superposition has ever been observed. If we add another system (even a cat or a man) to the initial system and the apparatus, it becomes entangled as well with the first two. This is the core of the celebrated Schrödinger's cat argument [5] and Wigner's friend problem [6].

It is worth noticing that there is a difference in the way it is possible to use these two rules. The Schrödinger equation can be used without any further knowledge on the system. All that is necessary to compute the future state at an arbitrary time is the initial state $|\Psi\rangle$ and the Hamiltonian H of the system. Using the reduction postulate requires the knowledge of the result provided by the measurement. That could seem harmless as it is the case for example in classical statistical physics where one has only a probability distribution on the possible states of the system and where one updates the state when learning in which one of the possible states the system really is. But in quantum physics, this statistical interpretation is totally ruled out as we are going to show.

2.2. Is a direct statistical interpretation possible?

It could indeed be tempting to think that quantum formalism is usable only for sets of systems and that an "easy" interpretation of superposed states is that they represent statistical descriptions of mixtures of systems each one in a definite state corresponding to a classical state.

In this case, a set of N systems in the state $|\Psi_S\rangle = \sum c_i |\varphi_i\rangle$ would be identical to a set of N systems such that a proportion $|c_i|^2$ is in the state $|\varphi_i\rangle$. Hence we could recover an interpretation similar to the interpretation of classical statistical physics. A first easy way to see that it is wrong is to consider the example of a set of N spin ½ particles in the state:

$$|\Psi\rangle = \frac{1}{\sqrt{2}} |+\rangle_z + \frac{1}{\sqrt{2}} |-\rangle_z \quad (4)$$

Let's call it set 1. A measurement of the spin along Oz of one of these particles will give the result + in half of the cases (and the result – in the other half). Assuming that this set 1 is identical to a set where half of the particles are in the state $|+\rangle_z$ (hence having a spin + along Oz) and the other half in the state $|-\rangle_z$ (hence having a spin – along Oz) is coherent with any measurement along the axis Oz (let's call this new set, set 2). But, if we measure the spin along the axis Ox, then the results are different. For set 1 the result will always be +, while for set 2, the result will be + in half of the cases and – in the other half. These two sets leading to different predictions for the measurement along Ox cannot be identical.

More generally, the interferences appearing between different terms of the superposition show that this interpretation is definitely wrong. Assume that we consider two observables A and B that do not commute and that the system is in a state:

---

[3] Actually, the apparatus being in an entangled state with the system has no state by itself strictly speaking since through this entanglement only the system S+A has a state. So, it is only a convenient way to speak to say that it is in a superposition of $|A_i\rangle$. The reason why we can safely do that is the fact that the predictions on the apparatus alone (considered as a sub system of S+A) that we can draw from the entangled state of S+A are similar to the predictions we could draw from an equivalent superposition of $|A_i\rangle$. The correct formalism in this case is the density matrix that we will describe in the following.

$$|\Psi_S\rangle = \sum_i \lambda_i |i_A\rangle = \sum_j \mu_j |j_B\rangle \quad (5)$$

with $\sum_i |\lambda_i|^2 = \sum_j |\mu_j|^2 = 1$ where the $i_A$ (resp. $j_B$) are the eingenvalues of A (resp. B) and the $|i_A\rangle$ (resp. $|j_B\rangle$) are the eigenvectors of A (resp. B) linked to the corresponding eigenvalues. Let's assume that:

$\langle i_A | j_B \rangle = v_{ij}$ so that $|j_B\rangle = \sum_i v_{ij} |i_A\rangle$. Hence:

$$|\Psi_S\rangle = \sum_j \mu_j |j_B\rangle = \sum_j \sum_i \mu_j v_{ij} |i_A\rangle \quad (6)$$

Let's consider a set of N systems such that a proportion $|\mu_j|^2$ is in the state $|j_B\rangle$. The probability to get the result $\lambda_i$ if a measurement of A is made on a system in the state $|j_B\rangle$ is $|v_{ij}|^2$ so the probability to get the result $\lambda_i$ if a measurement is made on a system from this set is $|\mu_j|^2 |v_{ij}|^2$. Now the probability to get the same result if a measurement of A is made on a system in the state $|\Psi_S\rangle = \sum_j \mu_j |j_B\rangle$ is $\left|\sum_j \mu_j v_{ij}\right|^2$ which is different from $|\mu_j|^2 |v_{ij}|^2$ because it contains the interference terms $\sum_{j \neq j'} \mu_{j'}^* v_{ij'}^* \mu_j v_{ij}$.

That proves that it is not possible to identify the superposed state $|\Psi_S\rangle = \sum_j \mu_j |j_B\rangle$ of a system to a state where the system is in one definite eigenvector $|j_B\rangle$ with the probability $|\mu_j|^2$ or to identify a set of N systems in the state $|\Psi_S\rangle = \sum_j \mu_j |j_B\rangle$ to a set of N systems such that a proportion $|\mu_j|^2$ is in the state $|j_B\rangle$. A superposed state is not interpretable as a classical description of the probabilistic ignorance about the real definite state in which the system is. Hence, if we consider that quantum mechanics is complete (i.e. that is cannot be completed by hidden variables as we will analyze in §3.1.2), a quantum measurement cannot be considered as a process simply revealing a pre-existing value.

## 2.3. What is a measurement?

The conclusion of §2.2 shows that the problem of the incompatibility between the two descriptions of a measurement remains unsolved. The problem is that both descriptions seem equally valid though they lead to totally different states. Now, the Schrödinger equation describes a linear and unitary process while the reduction postulate is neither linear nor unitary. That means that there is a priori no way we can get a reduction of the state through the Schrödinger equation.

That would not be a problem if it was possible to give a clear and not ambiguous definition of what a measurement is. In this case, we would have two well separated situations, a first one when no measurement is made on the system and a second one when a measurement is made. In the first case, we should apply the Schrödinger equation and in the second one, the reduction postulate.

Is it possible to define clearly what a measurement is? What many physicists are looking for is a definition that could be regarded as "strongly objective" in the meaning that d'Espagnat gave to this term [7] (i.e. without any mention to a human observer)[4]. Widely accepted by a large majority of physicists who wanted to believe that it solves all problems, the Copenhagen interpretation, mainly proposed by Bohr, says that a measurement is an interaction between the system and a macroscopic classical apparatus[5]. Is this interpretation strongly objective? In the following, we are going to see that, despite the appearance, it is not the case. We will also present and analyze several other interpretations that have been proposed to solve the measurement problem.

---

[4] See for example [8] for a typical exposition of this quest.

[5] It even goes as far as saying that the property that is measured belongs not to the system itself but to the whole composed of the system plus the apparatus.

### 3. Many interpretations

Faced to what seems a real inconsistency inside the quantum formalism, physicists have proposed many solutions largely depending on their initial philosophical inclination. A very rough description of the different families of positions is the following:

- A first one is the family of instrumentalist positions which consider that quantum mechanics is nothing but a tool to predict the results of the experiments done by the physicists. The formalism is then a list of recipes that must be applied in strictly following the rules and that do not speak of something else than the results of the experiments. For the instrumentalists, wondering if there is an external reality independent of any observer and trying to know the properties of this reality, or asking for what happens between the experiments, are at best questions they do not want to consider and at worst questions that they consider as meaningless. Bohr's position and the Copenhagen interpretation are instrumentalist as we will see. Belonging to the same family, the pragmatist interpretation (in the spirit of William James and John Dewey) tries to stick to the practice of the physicists and often takes as basic given facts some features (for example, the unicity of the result of a measurement) that other interpretations want to explain [9 - 11]. Some pragmatist positions (for example Healey [12]) are nevertheless not instrumentalist but partially realist.

- A second one is the realism (the association of the metaphysical realism, the epistemic realism and the scientific realism to be more precise) which considers that there is an external reality which exists independently of any observer (metaphysical realism), that this reality is roughly similar to the way it appears to us, that it would be unchanged even if there were no human beings, that it is possible to describe and to understand it (epistemic realism) and that the scientific formalism aims at describing the world as it really is (scientific realism). This was the natural position of the majority of scientists during the nineteenth century and it is in this framework that the standard initial formulation of quantum mechanics was done by Dirac and von Neumann. This was the position of Einstein, Schrödinger, Bell, de Broglie and Bohm. It is worth mentioning that despite the problems it faces, it is still the current philosophical position of many physicists who do not want to be involved in what they call "metaphysical questions".

- A third one is the idealism which gives the primacy to the spirit and considers that what we perceive is nothing but a creation of our mind. Many different sorts of idealism exist depending on the degree to which they accept or refuse the concept of reality. The most extreme kind of idealism is the solipsism which denies that there exist something else than one's own personal mind (as we will see, Convivial Solipsism is not that extreme and allows for the existence of other minds and of something external to the mind). More prudently, the idealism can limit itself to affirm that the only reality we have access to is our perceptions and that postulating every other kind of reality is at best risky. Kant's transcendental idealism does not deny that there is a reality, the things in themselves, but states that there is an unbridgeable gap between the phenomena that we perceive and the noumena that are unknowable and out of reach.

I am now going to present the main interpretations that have been put forward to try solving the measurement problem. First I will present solutions that imply modifying the quantum formalism and then I will stay for the remaining part of the paper inside the standard quantum mechanics.

### 3.1. Modifying quantum mechanics

#### 3.1.1. The "Correctness of the Quantum Formalism" assumption

I first start by making an assumption which seems very reasonable about the quantum formalism. Due to the huge number of successful tests that it has undergone, I will assume that all its presently testable predictions are right. That means neither that it is impossible that some prediction that is presently out of reach to test could not be false (we'll see below that some other formalisms can differ from the quantum one with respect to some predictions that are presently impossible to test) nor that quantum formalism is the only one able to make these predictions. It is even possible that another formalism makes some predictions not done by the quantum formalism provided that these predictions not be in disagreement with some prediction from the quantum formalism. This assumption only means that all the predictions from the quantum formalism that we could test in principle by an experiment that is currently feasible would be confirmed if we did the experiment. Hence we will demand from any rival formalism to make exactly the same predictions on pain of being disqualified.

Assumption of Correctness of Quantum Formalism (CQFA):

"*All the predictions made by the quantum formalism that are presently testable through a feasible experiment would be confirmed if we did the experiment*".

In this paper, I will suppose that CQFA is true and I will consider only alternative theories respecting CQFA.

#### 3.1.2. Welcoming hidden variables

Let's recall the conclusion of §2.2: solving the measurement problem in considering that the superposed state $|\Psi_S\rangle = \sum_j \mu_j |j_B\rangle$ of a system is a state where the system is in one definite eigenvector $|j_B\rangle$ with the probability $|\mu_j|^2$ or that a set of N systems in the state $|\Psi_S\rangle = \sum_j \mu_j |j_B\rangle$ is a set such that a proportion $|\mu_j|^2$ of the N systems is in the state $|j_B\rangle$ is not possible. That means that the pure quantum formalism does not allow to interpret the probabilistic predictions of a superposed state in a statistical way.

Is it nevertheless possible to consider that, even though quantum formalism is not suited to take them into account, properties of individual systems possess well defined values prior to the measurement that reveals them? These properties could be given by additional parameters completing the usual quantum description in such a way that, though the quantum state does not allow to predict the value possessed by an observable, this observable has nevertheless a definite value that would be computable in principle, would one know the unknown parameters linked to the system. These parameters are called hidden variables. The state vector would then describe an ensemble of systems and each individual element of this ensemble would have a definite value for every observable that would be revealed by the measurement. In this case, we could recover a statistical interpretation of the probabilistic predictions of quantum mechanics. This is the hope of those, as Einstein, who think that quantum mechanics is not complete.

For several decades, this hope seemed impossible due to a proof given by von Neumann in 1932 [4] that hidden variables were impossible. It is now well known that von Neumann's proof relies on an assumption that is much too demanding. To understand that, let's consider a reasonable set of assumptions for a hidden variables theory[6]. Given a state representing an ensemble of systems with observables A, B, C, …, to each system is assigned a numerical value v(A), v(B), v(C), …, for each observable, in such a way that if a measurement is performed on a set of commuting observables, the results must be the corresponding values. The theory must give a rule for every state explaining the distribution of these values over the elements of the ensemble in order to recover statistically the predictions of quantum mechanics when many measurements are performed on the systems of this ensemble. Of course, the allowed values for an observable are limited to its eigenvalues. Moreover, since any relation f(A,B,C,…) = 0 satisfied by a set of commuting observables is also satisfied by the eigenvalues linked to their common eigenvectors, it has also to be satisfied by the values assigned to any individual system, so f(v(A),v(B),v(C),…) = 0. In particular, if A and B commute, v(A+B)=v(A)+v(B)[7]. Von Neumann's assumption was to require that last relation to hold even when A and B do not commute[8]. Having rejected this condition, Bell constructed a very simple hidden variables model for one spin ½ fully satisfying the more reasonable requirements above [14]. Would it be possible to generalize this model, we would have a possible solution to the measurement problem since the superposed states of quantum mechanics could merely be interpreted as statistical sets of systems having well defined properties. A hidden variables theory would be in the same relation to quantum mechanics than classical mechanics is to classical statistical mechanics. But things are not so simple. What has been possible for one single spin ½ particle is no more possible for two independent spin ½ particles. Let's consider the Pauli matrices $\sigma_\mu^1$ and $\sigma_\nu^2$ and the nine observables shown in the figure below:

$$\sigma_x^1 \quad \sigma_x^2 \quad \sigma_x^1\sigma_x^2$$

$$\sigma_y^2 \quad \sigma_y^1 \quad \sigma_y^1\sigma_y^2$$

$$\sigma_x^1\sigma_y^2 \quad \sigma_x^2\sigma_y^1 \quad \sigma_z^1\sigma_z^2$$

---
[6] I follow here Mermin [13] for a very simple proof of the Bell, Kochen, Specker theorem.

Fig 1.

Let's prove that it is not possible to assign values to all these nine observables. It's easy to check that the observables in each row and each column are mutually commuting. The product of the three observables of the right column is -1 while the product of the three observables of the other columns as well as the product of the three observables of each row is +1. As we have seen, the values assigned to mutually commuting observables must obey any identities satisfied by the observables themselves. But this is impossible since the product of the values of the nine observables computed as the product of row 1 with row 2 and row 3 is +1 while the same product computed with the columns is -1.

An important remark is that in this proof it is assumed that each observable has a value in an individual system independently of how that value is measured and of which other commuting observables are measured simultaneously. This condition is referred to as noncontextuality. So this is a proof that a hidden variables theory in agreement with quantum mechanics cannot be noncontextual[9].

---

[7] Indeed, if A and B commute, the observable C = A+B commute also with A and B and from C-A-B=0 one draws v(C)-v(A)-v(B)=0.

[8] Von Neumann assumed this condition because quantum mechanics requires that for any state $|\Psi\rangle$, $\langle\Psi|A + B|\Psi\rangle = \langle\Psi|A|\Psi\rangle + \langle\Psi|B|\Psi\rangle$. But of course, when A and B do not commute, this is true only in the average. So von Neumann's assumption means demanding that a relation holds in the mean by imposing it case by case. An assumption that Bell even qualified as silly.

We will see later that another Bell theorem proves that any theory reproducing the results of quantum mechanics must be non-local. Hence, a successful hidden variables theory must be contextual and non-local. The most famous example of such a theory is the Bohm theory [16] in which the measurement problem does not exist. Bohm theory fully respects CQFA. Bohm postulates beside the usual wave function an actual configuration through the positions $q_i$ of the particles which evolves through a guiding equation:

$$\frac{dq_i}{dt} = \frac{\hbar}{m_i} Im \frac{\Psi^* \nabla_i \Psi}{\Psi^* \Psi} \qquad (7)$$

At any moment the particles have a definite position that exists even when unobserved and they follow a definite trajectory. The evolution of the wave function is still given by the Schrödinger equation. In order to reproduce the usual quantum predictions, the initial distribution of positions must satisfy the Born rule given by $|\Psi|^2$. It is possible to show that, due to the dynamic described by the guiding equation, when a particle interacts with a macroscopic apparatus, the latter is in a unique position at the end of the interaction. Indeed, the global wave function evolves into a superposition of states that remain separated in the configuration space. Hence the configuration stays inside a unique component of the wave function, as if a collapse had occurred. There is no more need for a reduction postulate and the measurement problem is dissolved. This is undoubtedly a great success of this theory.

### 3.1.3. Modifying Schrödinger equation

The Schrödinger equation being linear and describing a deterministic and unitary evolution cannot give a probabilistic reduction of the state vector. Hence it is natural to try modifying the Schrödinger equation in a way that allows getting a reduction when a measurement is done but preserving of course the current predictions when no measurement is made. The most famous attempt in this direction has been done by Ghirardi, Rimini and Weber (GRW theory) [17] who add stochastic and nonlinear terms in the Schrödinger equation. The wave function is subjected at random times to a spontaneous localization in space and the frequency of this localization increases with the number of constituents of the system. When the parameters of this process are well tuned, this allows macroscopic systems to be well localized without changing the usual quantum predictions for micro systems. Actually this is not exactly true since GRW theory makes some predictions that disagree with the predictions of quantum mechanics even though only in a presently non-testable manner [18]. This is the reason why GRW theory, even contradicting quantum mechanics on some points, is not eliminated by invoking CQFA. Perhaps future experiments will decide in favor of GRW theory or will falsify it. Nevertheless what is interesting is that, at least in principle, GRW theory is a rival theory for quantum mechanics.

### 3.1.4. Are these modifications acceptable?

We will not discuss here in details if the modifications of quantum mechanics through GRW theory or Bohm theory are acceptable or not and what is the price to pay for switching from the standard quantum theory to these other theories. Let's just say first that building relativistic versions of these theories is difficult even if recent developments seem to indicate that it is possible in both cases[10]. Moreover each one of these theories suffers from some unsatisfactory points. GRW theory explains the reduction through a mechanism that seems truly ad hoc. What is the reason for these spontaneous random localizations? There is none. They must be accepted as they are. Moreover, the parameters of the random localizations are tuned by hand in order to fit with the microscopic observations. This is of course not enough to reject the theory but this is not very satisfying. Bohm theory was initially intended to restore the intuitive representation of the particles and their trajectories whereas quantum mechanics forbids to speak about them. But it appears that the picture that is given by the theory is far from allowing to come back to a simple realism similar to the one we can assume in classical physics. The first point is of course that the theory is contextual and non-local, but this is unavoidable due to BKS theorem and to the violation of Bell's inequalities (as will be discussed In §5). That means that the value of an observable cannot be considered as belonging directly to the system itself but depends also on the experimental set up that is used to measure it. In particular, measuring an observable A along with the two observables B and C (where A, B and C are mutually commuting) will not yield the same value for A than measuring it along with two other observables B' and C' (where A, B' and C' are also mutually commuting). Even worst, the value measured for A on one particle will possibly depend instantaneously on what is measured for B on another particle even if the second measurement is made very far from the first one. These features are unavoidable and common to every theory that preserves the predictions of quantum mechanics. So, it is hard to blame Bohm theory for that but it removes a large part of its initial appeal. Another point is that even if each particle follows a trajectory, this trajectory is unobservable since the very fact of observing it requires to use an experimental set up which changes the quantum potential which guides the particle and hence the trajectory. This is totally different from the case of classical trajectories which are observable without modifying them, at least in principle. So Bohm theory fails to help us recovering an intuitive realist vision which nevertheless was its initial motivation. Moreover the fact that the initial distribution of positions must satisfy the Born rule given by $|\Psi|^2$ can be considered as a very strong assumption.

---

[9] The usual presentation of contextuality is given through the observables of a spin 1 particle. The observable $S^2$ which is the sum of the square of the components of the spin along any three orthogonal directions is equal to 2. Since the unsquared components have eigenvalues equal to -1, 0 or 1, that implies that one must be equal to 0. It is then possible to show that it is impossible to assign a value to one spin component without deciding which the orthogonal directions are [15].

[10] See [19] for Bohm theory and [20, 21] for GRW theory.

The question of deciding if either of these theories should be accepted is left open[11]. In the following of this paper, I will assume that we fully accept the standard quantum formalism as a framework for the discussion[12].

### 3.2. Bohr's position and the Copenhagen interpretation

The Copenhagen interpretation is mainly due to Bohr[13] with major contributions from Heisenberg[14] and Born. Before the article[15] of Einstein, Podolski, Rosen in 1935 [24] (hereafter EPR paper), Bohr's view was that observing a quantum object involves an uncontrollable physical interaction with a classical measuring device that affects both systems. This interaction produces an uncontrollable effect and that is the reason why the result of the measurement can only be predicted statistically. This effect cannot be neglected because the quantum of action is not null. When the position of an electron is measured, the apparatus and the electron interact in an uncontrollable way, and it becomes impossible to measure the momentum of the electron at the same time. A pictorial representation of this idea was given by Heisenberg in a thought experiment called Heisenberg's microscope [25]. That is the essence of the concept of complementarity: the measurement of the position of a system affects its momentum uncontrollably. Thus it is not possible to determine precisely both position and momentum.

---

[11] See [22] for an open discussion between several physicists on the reasons to accept or to reject Bohm theory.

[12] Of course, this choice is not neutral for the philosophical discussion and we have to make clear that the conclusions that we will get could be put in question in the case where either of these theories are proved to escape all the objections that today prevent from considering that it is the "best" theory according to its empirical predictions, to its fruitfulness and to its adequacy with our preexisting philosophical requirements (which is probably the most problematic aspect).

[13] See [23] for an extended description of Bohr's position.

[14] Actually, Bohr and Heisenberg were not in total agreement. Heisenberg's position was considered as too subjective by Bohr.

[15] See below §4 for a detailed presentation of the EPR Paradox.

Complementarity helps understanding the Heisenberg uncertainty relations and is also the source of the statistical character of the quantum predictions. After the EPR paper, Bohr's position evolved into a more radical departure from the vision that microscopic systems were classical objects possessing definite properties (such as it is depicted in Heisenberg's microscope experiment). He started to say that a measurement is an interaction between the system and a macroscopic classical apparatus and that the property that is measured belongs not to the system itself but to the whole composed of the system plus the apparatus. He even went so far as saying that the very meaning of a property depends on the experimental set up that is used and that speaking about a property of a system without making reference to the apparatus that is used for measuring this property is meaningless. In the following, we will focus on this mature version of Bohr's ideas which is often considered by physicists not willing to be involved in philosophical difficulties to eliminate all the problems of interpretation of quantum mechanics. For them, since a physicist using a macroscopic apparatus for measuring a physical property on a system perfectly knows what he is doing, there is no ambiguity and the reduction postulate must be applied. So, there is no more problem. Even if this point of view is perfectly working in practice from an instrumentalist standpoint, it leaves open the question of knowing what a macroscopic apparatus is. The distinction micro / macro is not a sharp one and assuming naively that a macroscopic system behaves classically is problematic since we know of many macroscopic systems showing a quantum behavior (super conductivity, super fluidity etc.). Another difficulty with this view is that the quantum formalism is assumed to be universally valid and it should be applied to any physical system, whether microscopic or macroscopic. But actually a more careful analysis of Bohr's position shows nevertheless that what he had in mind was not that the classical behavior of the apparatus was directly linked to its macroscopic aspect but that it was linked to the use that the observer wanted to make of it. Bohr had in mind pragmatic reasons. If the observer wants to make a measurement on the apparatus through another apparatus, it has a quantum behavior. If it is used to make an observation on another system, it has a classical behavior [7, 23, 26]. In this case, it is clear that the role of a human observer in the definition of the measurement process cannot be avoided and that this definition can hardly be considered as "strongly objective" in the sense that we mentioned in §2.3. What is clear is that Bohr never explained precisely what a measurement is and left open the question of determining when a measurement happens except for the case when a human observer knows that he is doing one. But he never explicitly gave this definition and avoided to be involved in mentioning the concept of consciousness. So, even if it contains many profound features, Bohr's position can hardly be considered as the definitive solution to the measurement problem.

3.3. Interpretation involving the consciousness of the observer

Faced to the difficulty to avoid mentioning consciousness during the measurement process, some authors considered that even if the quantum formalism has a physical universal validity, the consciousness of the observer lies outside of its scope. As von Neumann says [4]:

*"Experience only asserts something like: an observer has made a certain (subjective) perception, but never such as: a certain physical quantity has a certain value."*

*"It is inherently entirely correct that the measurement or the related process of the subjective perception is a new entity relative to the physical environment and is not reducible to the latter."*

*"But in any case, no matter how far we calculate, [….], at some time we must say: and this is perceived by the observer. That is, we must always divide the world into two parts, the one being the observed system, the other the observer."*

Wigner adds [27]:

*"When is the measurement completed? We will see that […] the measurement is completed only when we have observed its outcome"*

More precisely, according to their view, a measurement is made only when the consciousness of an observer has an interaction with a system. This interaction between the system and the consciousness of the observer has the physical effect to change the state of the system and to project it according to the reduction postulate. This is the position stated by Wigner [28] and by London and Bauer [29]. As we will see, these authors got a correct intuition about the fact that consciousness cannot be eliminated from the measurement process and that it is impossible to give a strongly objective definition of what a measurement is in quantum mechanics. But from this correct premise they draw a wrong conclusion. Assuming that, lying outside of the field of the physical entities obeying quantum mechanics, consciousness has nevertheless a physical action on the state of the entities inside this field is altogether shocking and incoherent. Of course, their position received a wide opposition from a majority of physicists not satisfied by this renaissance of the old Cartesian dualism. In Convivial Solipsism, consciousness plays also an essential role but does not interact with systems in a physical sense.

### 3.4. Interpretations considering the vector states as relative

I present in the following three interpretations that share the feature that the vector states are no more absolute but relative. Of course, the way they are relative and to what they are relative differ from one interpretation to the other. But this is as well an important feature of Convivial Solipsism and I need to clarify the similarities and differences between the way this relativity is used in these interpretations and in Convivial Solipsism.

#### 3.4.1. Everett's interpretation

Everett was worried by the measurement problem that he was seeing as a fundamental inconsistency of the standard collapse formulation of the quantum mechanics given by von Neumann [30]. He noticed that in the Copenhagen interpretation of the standard theory, the observer must be treated as an external system and that this prevents the universe to be described as a whole. He proposed [31, 32] a relative-state formulation of pure wave mechanics without reduction. In his interpretation, there is no collapse and the universal wave function (observers included) evolves uniquely under the Schrödinger equation. Actually, there are several ways to understand Everett's interpretation[16] and I'll present here only Everett's own one and the one given by Graham and De Witt [33] usually called the "many-worlds interpretation" that is the way the majority of physicists understand Everett's interpretation even though it was not supported by Everett.

Everett was interested in giving a coherent explanation of the Wigner's friend problem [6].

Let's for example, consider the measurement by a friend of mine F of a system S with an apparatus A. Assume that S is a spin half particle in a superposed state along Oz.

$$|\Psi_S\rangle = \alpha \,|+\rangle_z + \beta \,|-\rangle_z \tag{8}$$

A is a Stern and Gerlach apparatus with a magnetic field oriented along 0z and F performs a measurement of the spin of S using A. After the interaction between my friend, the apparatus and the system, the Schrödinger equation gives the final entangled state[17] for the whole system + apparatus + friend:

$$|\Psi_{SAF}\rangle = \alpha \,|+\rangle_z |\uparrow\rangle |F_+\rangle + \beta \,|-\rangle_z |\downarrow\rangle |F_-\rangle \tag{9}$$

---

[16] In particular there is the so called "many-minds" version [34, 35] that I will not analyze here.

Where $|\uparrow\rangle$ and $|\downarrow\rangle$ are the states of the apparatus with impact up (for spin up) and down (for spin down) and $|F_+\rangle$ and $|F_-\rangle$ are the states of the consciousness of my friend having seen a result "up" or "down".

If I include myself in the global state I get:

$$|\Psi_{SAFO}\rangle = \alpha \,|+\rangle_z |\uparrow\rangle |F_+\rangle |☺\rangle + \beta \,|-\rangle_z |\downarrow\rangle |F_-\rangle |☹\rangle \tag{10}$$

Where $|☺\rangle$ and $|☹\rangle$ are my state of consciousness after having asked to my friend what she has seen and having received the answer "up" or "down".

Usually, the reduction postulates is used to insure that I will end either in the state $|☺\rangle$ or in the state $|☹\rangle$. But if there is no collapse as Everett assumes, it is necessary to explain why I see a definite value and not a superposition. For Everett, observers are automatically functioning machines possessing recording devices (memory). The result of a measurement made by an observer is the state of her memory. Everett then makes a crucial distinction between absolute and relative states. For him, subsystems do not possess states that are independent of the states of the remainder of the system. If in this case, $|\Psi_{SAFO}\rangle$ can be considered as absolute because it concerns the whole system, while $|☺\rangle$ and $|☹\rangle$ are relative because they concern only a part of the system.

*"Let one regard an observer as a subsystem of the composite system: observer + object-system. It is then an inescapable consequence that after the interaction has taken place there will not, generally, exist a single observer state. There will, however, be a superposition of the composite system states, each element of which contains a definite observer state and a definite relative object-system state. Furthermore, as we shall see, each of these relative object system states will be, approximately, the eigenstates of the observation corresponding to the value obtained by the observer which is described by the same element of the superposition. Thus, each element of the resulting superposition describes an observer who perceived a definite and generally different result, and to whom it appears that the object-system state has been transformed into the corresponding eigenstate."* [36].

So each relative memory state describes a relative observer with a determinate measurement. $|☺\rangle$ and $|☹\rangle$ are then relative memory states of my consciousness describing different results. Everett insists on the fact that this in full agreement with our experience because no observer can be aware of the branching when it occurs. Now, an important and often forgotten point of his position is that Everett considers that all elements of a superposition must be regarded as simultaneously existing. In particular, that means that he thinks that it would always be possible in principle to put other branches in evidence through a measurement that would show interferences between them since the universal state vector remains superposed. The picture is then a unique universe described by a universal entangled state vector whose dynamics is entirely described by the Schrödinger equation and observers entangled also with the rest of the universe but having memory records corresponding to their relative states that are associated with experience of a definite result. Each observer has several different memory records (as many as possible results) but remains unique.

The "many-worlds" interpretation (which is actually the most frequent way Everett's interpretation is presented) gives a different picture. Graham and De Witt give the following description [33]:

---

[17] In the following, we will often omit the symbol ⊗ for the tensorial product.

> *"By virtue of the temporal development of the dynamical variables the state vector decomposes naturally into orthogonal vectors, reflecting a continual splitting of the universe into a multitude of mutually unobservable but equally real worlds, in each of which every good measurement has yielded a definite result and in most of which the familiar statistical quantum laws hold."*

So, if a measurement with a probabilistic outcome is made, the world splits into several worlds and each possible outcome of this measurement appears in some of these worlds in a proportion given by the quantum probability of that outcome. The main difference with Everett's initial interpretation is that the universe is not made of a unique world but contains an exponentially growing number of worlds that are causally isolated and a corresponding growing number of clones of the initial observer each one inhabiting one world. No need to say that this version is not particularly economic from the ontological point of view. What is interesting is that it would be possible in principle to make a crucial experiment to decide between the two versions through an attempt to detect some interferences between branches, predicted by the first version and forbidden by the second one. The decoherence theory (see §7), correctly interpreted, works in favor of the initial version and not of the "many-worlds" one.

Actually Vaidman [37, 38] defends a "many worlds" interpretation which is very close to the initial Everett position and not subject to the same criticism than Graham and De Witt's one. The main difference is that in Everett's picture, worlds are relative to us and independent of our concepts while Vaidman prefers to define the concept of world relative to our concepts and independent from us, especially for helping discussion about times without observers (far past and far future).

However, in all these versions, Everett's interpretation faces two problems. The first one is the problem of the preferred basis (see below §7.3). But that could be solved by adding decoherence to it as we will see later. The second one is recovering the quantum statistics and Born rule in a sequence of measurement records. Attempts for that have been proposed but no general agreement has been reached[18]. Moreover it is not very satisfying to deal with so many consciousness for the same observer, which is the case in all versions.

Everett's interpretation can be read as a realist position even if the reality it describes is not very familiar. In both versions, the universe exists independently of any observer and is described by the universal state vector whose dynamics obeys the Schrödinger equation, is never reduced and remains entangled. Hence, this reality is not similar to the reality we are accustomed to. Observers are included in this universe as a part of it and their mental states are described by their relative states. What they see (which is not truly the reality since the universe remains in a superposed state) is however conform to the reality we are accustomed to.

### 3.4.2. The relational interpretation

The relational interpretation is due to Rovelli who denies that the state of a system is observer-independent but, on the contrary, says that, similarly to what is the case in special relativity, a system has a state only relative to a given observer [41 - 43]. Moreover, for him, an "observer" is not mandatorily a human being but can be any physical system. So Rovelli belongs to the family of those who want to get a strongly objective formulation of quantum mechanics without any need of the concept of consciousness. An observer provides a frame of reference relative to which states and values can be assigned much in the same way that the speed of a system is relative to an inertial reference frame. He tries also to derive the formalism of the theory from a set of assumptions based on the concept of information. For him, quantum mechanics is a theory about information. In the following I will not

---

[18] See for example Wallace [39] who claims having proved the Born rule in this context and Kent [40] who denies that it is the case.

address this second point[19] and will focus mainly on the question to know if assuming that the state of a system is observer-dependent solves the puzzles of quantum mechanics as he claims or if this is not the case. I must say first that I am very sympathetic with the view that states and values are relative to a given observer. As we will see in the last part of this paper, Convivial Solipsism shares this particular feature and extends it even further. That's the reason why, I will examine the relational interpretation in details. But, as we shall see, Convivial Solipsism departs from the relational interpretation in many respects mainly about what an observer and a measurement are. For Rovelli [41] any physical system can be an observer:

*"By using the word 'observer' I do not make any reference to conscious, animate, or computing, or in any other manner special, system. […] Velocity is a relational notion (… ) and thus it is always (explicitly or implicitly) referred to something; it is traditional to denote this something as the observer, but it is important to keep in mind that the observer can be a table lamp".*

A table lamp is a macroscopic system but this is not even necessary and Rovelli insists on the fact that nothing distinguishes macroscopic systems from quantum systems.

*"If there is any hope of understanding how a system may behave as observer without renouncing the postulate that all systems are equivalent, then the same kind of processes –"collapse"– that happens between an electron and a CERN machine, may also happen between an electron and another electron".*

Hence, an electron can play the role of an observer relative to another electron. In summary, from the "point of view" of any physical system (be it an electron) any interaction with another system is a measurement allowing the second system to possess some definite physical values relative to the first one, much in the same way that from the point of view of an inertial reference frame, a car has a definite speed.

Rovelli exposes the (usual since von Neumann) two ways of describing a measurement process similarly to the presentation given above in §2.1. He considers an observer O (a classical measuring apparatus not necessarily including a human being) that makes a measurement on a system S. Assuming that the physical property q being measured takes two values 1 and 2 and that S is in an initial state $|\Psi_S\rangle = \alpha|1\rangle + \beta|2\rangle$, O measures either one of the two values 1 and 2 with respective probabilities $|\alpha|^2$ and $|\beta|^2$. Assuming that the outcome of the measurement is 1, the initial S state is projected onto $|1\rangle$. Rovelli then considers the same sequence of events from the point of view of a second observer P who describes the interacting system formed by S and O. P knows the initial states of both S and O but does not perform any measurement on the system S-O. Let's say that O evolves to $|O_1\rangle$ (resp $|O_2\rangle$) when 1 (resp 2) is the outcome of the measurement made on S. Hence P assigns the state $|\Psi_{OS}\rangle = \alpha|1\rangle \otimes |O_1\rangle + \beta|2\rangle \otimes |O_2\rangle$ to the system S-O. Rovelli points to the fact that we have two descriptions of the same sequence of events, one given by the observer O, in which S is in the state $|1\rangle$ and one given by the observer P, in which S is not in the state $|1\rangle$. Usually the fact that these two descriptions are incompatible is considered as a problem (see §2.1) but Rovelli says that there is no problem since both descriptions are correct. There is no unique true description because according to him:

*"In quantum mechanics different observers may give different accounts of the same sequence of events".*

---

[19] This point is interesting for the attempt to derive quantum formalism from simple assumptions but adds nothing to the discussion about the measurement problem since Rovelli's solution for the puzzling problems of quantum mechanics is mainly contained inside this idea of observer-dependent state.

Is this position really solving the problem? The first objection is that Rovelli assumes that a measurement of q is made by O on S and that q has (for O) a definite value, say 1. But he doesn't explain why the initial superposed state of S is projected onto one eigenvector of q. If, this is because a measurement has been performed on S by O, it is necessary to explain what a measurement is, what Rovelli does not. If O can be any physical system (what he claims) then it is difficult to understand why a simple interaction between O and S will have the result of projecting for O the superposed state of S onto one eigenvector, unless using a new ad hoc postulate, at least as embarrassing than the usual reduction postulate. It seems as if Rovelli postulated that any interaction between two physical systems (usually described as an entanglement between the two systems) has the consequence that from the point of view of the first one, the state of the second one is reduced (and vice versa). That seems strange to say the least. This postulate is much stronger than the usual reduction postulate. Of course, if we accept it, then the reduction postulate becomes a mere consequence. But why is the unitary evolution broken down (from the point of view of each system) during an interaction? According to Rovelli, the reason is because O does not have a full dynamical description of the interaction:

*"O cannot have a full description of the interaction of S with himself (O), because his information is correlation, and there is no meaning in being correlated with oneself"*.

Even if it is possible to concede that O does not know the Hamiltonian of interaction between O and S (but one can wonder what is the meaning of the sentence "O does not know something" when O is an electron!), that does not explain why a single definite outcome is selected for O during the interaction. Now, if Rovelli accepts the usual reduction postulate when a measurement is made, he must explain why a simple interaction between any two physical systems is a measurement.

Moreover sometimes, Rovelli seems to confuse the fact a) that, following the interaction between S and O, there is a correlation between the state of S and the state of O (if S is in the state $|1\rangle$ then O is in the state $|O_1\rangle$ and if S is in the state $|2\rangle$ then O is in the state $|O_2\rangle$), with the fact b) that a measurement has been made (which means that S and O are in a definite eigenstate):

*"The fact that the pointer variable in O has information about S (has measured q) is expressed by the existence of a correlation between the q variable of S and the pointer variable of O. The existence of this correlation is a measurable property of the S-O state"*. [41]

But this correlation reflects merely the entanglement between the two systems while a measurement implies a reduction to only one of the two possibilities: either $|O_1\rangle|1\rangle$ or $|O_2\rangle|2\rangle$. Assimilating a correlation to a measurement seems difficult to accept[20]. And if it is not the case, one does not see why every interaction between two systems (even microscopic ones) should result in a measurement selecting one of the possible values.

Besides that, the famous problem of the preferred basis (see §7.3) is not mentioned by Rovelli. Implicitly he assumes that the observer to which the state of the system is relative is the one that is correlated to the "good values" corresponding to macroscopic not superposed states. His example of a measuring apparatus is misleading because it contains the implicit assumption that the only results that can be obtained are the eigenvalues of the observable that is measured. So far so good when the observer is a measuring apparatus designed to measure a specific observable. But Rovelli claims that the same

---

[20] Rovelli claims nevertheless that the assumption that an observer is merely a physical system having got information (i.e. correlation) on another system lies at the heart of the relational interpretation [22].

"measurement" situation can occur between two electrons. In this case, the second electron initially in a superposed state acquires a definite state relative to the first one. But what does that mean? Let's consider two electrons U and V such that from the point of view of the observer P who does not make any measurement on the pair, they are in the singlet state. What is the state of the second electron relative to the first one? Which observable has been measured? The state of the pair relative to P is:

$$|\psi_P\rangle = \frac{1}{\sqrt{2}}\left[\,|+\rangle_z^U\,|-\rangle_z^V - |-\rangle_z^U\,|+\rangle_z^V\,\right] \qquad (11)$$

According to the relational interpretation, the state of V relative to U can be either $|-\rangle_z^V$ or $|+\rangle_z^V$. But the singlet state is rotationally invariant and can be written under the same form whatever the direction of spin is chosen. So the state of V relative to U could also be $|-\rangle_x^V$ or $|+\rangle_x^V$. This is left unspecified by the relational interpretation.

Rovelli insists also on the fact that the comparison between different views (of different observers) is always a physical interaction which is quantum mechanical in its essence.

*"Two observers can compare their information (their measurement outcomes) only by physically interacting with each other".*

I fully agree with that. This is an essential feature that is shared by Convivial Solipsism. That is the reason why the consistency between the measurements of two different observers is preserved. If P asks a question to S concerning q and asks the same question to O (which can be done only by a measurement on S and a measurement on O), she will get consistent results. As Rovelli says [42]:

*"More precisely: everybody hears everybody else stating that they see the same elephant they see."*

Asking if O and P have "really" the same information about S is meaningless because this question would have a meaning only relative to a third observer and as Bitbol [44] correctly notices, there is no meta-observer able to witness neutrally what the other observers "really" see, so there is no way to compare what they have seen in the absolute. Hence it is important to stress the following point: does the agreement between O and P when P asks to O which result she has got, mean that if O has seen 1, then P will get the answer 1? What we have in mind if we think that it should be the case, is that at the end, the system S will be in the state $|1\rangle$, O in the state $|O_1\rangle$ and P in the state $|P_1\rangle$. In Rovelli's words, if I hear you telling me that you see an elephant, then you really see an elephant and not a tiger.

But this is not true for two reasons: the first one is that accordingly to the relational interpretation, we should not forget that there is no meaning speaking of a state without mentioning the observer to which it is related. So the correct way to express this idea inside the relational interpretation is that after the interaction between O and S, S is in the state $|1\rangle$ relative to O and S+O is in the state $|\Psi_{OS}\rangle = \alpha|1\rangle|O_1\rangle + \beta|2\rangle|O_2\rangle$ relative to P. Then P asks the question to O. Now according to the state $|\Psi_{OS}\rangle$, P can get either the answer 1 or the answer 2. Let's assume that she gets the answer 2. Then for her, the state of S+O becomes $|2\rangle|O_2\rangle$. Hence if she measures S, she will of course find a consistent result and will find the system S in the state $|2\rangle$ even if for O, S is in the state $|1\rangle$. That is the real meaning of the fact that there can be no disagreement between O and P.

The second reason is the point noticed above, that this sort of description is given from the point of view of a meta-observer which does not exist (a standpoint from nowhere) and so this way of understanding the situation is misleading. The fact is that there is no meaning of speaking in the absolute of what O and P have seen separately and to be surprised that they may have not seen the same thing. The accounts from O and P should not be juxtaposed. This point has been clearly developed by Van Fraassen [45] and Brown [46]. This is an unavoidable feature that will also be the case in Convivial Solipsism.

Einstein's standpoint was a realist one. Relativity is clearly compatible with the existence of an external independent reality and the relational interpretation is also compatible with an external reality even though this reality is relative to each system.

The main problem with the relational interpretation is that Rovelli assumes that any interaction between any two physical systems (even microscopic ones) is a measurement leading to the collapse of the state of the second one relatively to the first one (and vice versa). So doing, he thinks eliminating the need to include mind or consciousness inside his ontology. For him, we, as any other system, see a definite value when we have interacted with a system in such a way that a correlation has occurred between some eigenstates of one observable of the system and some definite states of us. But which states? Eigenstates of our own pointer observable? We have seen that the way the collapse occurs is not fully determined (e.g. the example of a singlet state). That faces the problem of the preferred basis that we'll examine in §7.3. Moreover, accepting that such a collapse occurs at any interaction giving birth to correlations is difficult to swallow and is even more puzzling than accepting the usual reduction postulate in case of a measurement. Even if this collapse is relative to the particular system having interacted and not absolute, does it occur really in the external reality of this particular system? If one of the systems is a conscious observer doing a measurement on a particle, does this collapse occur really (in the reality relative to the observer) for the particle that is measured and does it change its physical state? If not, what does that mean? If yes, what is the physical process leading to this projection of a superposed state onto a definite state? Rovelli, faithful to the way Einstein showed that properties such that positions and speeds are relative to a framework and willing to extend that to quantum properties, could reply that in special relativity, the speed of a system has no definite value if we do not specify a framework. He could add that, relative to one framework, the speed acquires one definite value without any need of a physical process explaining this "reduction". Hence he could argue that the fact that the spin of an electron is determined relative to a system once we have specified this system (having interacted with the electron) is similar and does not require any further explanation. But this is unacceptable since the situation is totally different in quantum physics: for example, the fact that a system has a speed that is undefined if no framework is specified does not lead to any interference between the "possible speeds" while in quantum mechanics there are interferences which could be observed between the different possible states of spin. The indefiniteness of the speed in the absence of a specified framework is not at all comparable to the superposition of undefined quantum properties. That is the limit of the parallel that Rovelli wants to use between special relativity and quantum mechanics.

Any attempt to give an interpretation free from the concept of conscious observer must nevertheless be able to explain why we (conscious observers) see what we see. Classical physics does that through the implicit assumption of a physical realism having the trivial consequence that if the world is "like that" independently of us then we see it "like that". As we have seen, quantum mechanics forbids to make such an assumption. Hence, it is necessary for any interpretation to explain how what it describes results in what we see. The entangled state in which we are (with the system, the apparatus and the environment) provides indeed correlations but is not interpretable as a state in which we see definite values for speed or position (unless we adopt the Everett interpretation). So, even if the reality is not absolute but relative to each of us, why do we see our own reality as if it was classical? Rovelli's interpretation has nothing to answer.

We will see later that even if it cannot be considered as providing a complete solution to the measurement problem, the theory of decoherence gives a satisfying explanation of why we never see macroscopic superpositions and why the pointer basis is emerging as the preferred basis. But Rovelli does not appeal to this theory.

### 3.4.3. The Quantum Bayesianism: QBism

QBism[21] is mainly supported by Caves, Fuchs, Schack and Mermin [47 - 50]. QBists think that the primitive concept of experience is the central subject of science. This seems similar to Bohr's position who insisted on saying that experiments were the basic subject of quantum mechanics but what Bohr had in mind was the classical reading of a macroscopic apparatus while for QBism, "experience" is much more than that. The word "experience" must be understood here as "personal experience", meaning for each agent what the world has induced in her throughout the course of her life. Experience is the way in which the world impinges on any agent and how the agent impinges on the world. It is partly an instrumentalist position since, for QBists, quantum mechanics is but a tool allowing any agent to compute her probabilistic expectations for her future experience from the knowledge of the results of her past experience. QBists adopt the subjective interpretation of probability[22] according to which probabilities are representing the degree of belief of an agent and, hence, are particular to that agent. This is an important point to notice: as they say, it is a "single user theory". That means that:

*"Probability assignments express the beliefs of the agent who makes them, and refer to that same agent's expectations for her subsequent experiences"* [50].

QBism shares with the relational interpretation and Convivial Solipsism the feature that the entities of the formalism (wave functions, probabilities) have no absolute value (are not necessarily the same for all the observers) but are relative to a particular system (a particular agent for QBism, a particular physical system for the relational interpretation, a particular conscious observer for Convivial Solipsism). QBists refuse the idea that the quantum state of a system is an objective property of that system. It is only a tool for assigning a subjective probability to the agent's future experience. So quantum mechanics does not directly say something about the "external world".

*"But quantum mechanics itself does not deal directly with the objective world; it deals with the experiences of that objective world that belong to whatever particular agent is making use of the quantum theory."* [50].

A measurement (in the usual sense) is just a special case of what QBism calls experience and that is any action done by any agent on her external world. The result of the measurement is the experience that the agent gets from her action on her personal world. In this sense, measurements are made continuously by every agent. A measurement does not reveal a pre-existing state of affairs but creates a result for the agent. The goal of quantum formalism is only to give recipes to allow agents to compute their personal degree of belief about what will happen if they do such or such experience. The way QBists solve the measurement problem is very simple: they assume from the beginning that the direct internal awareness of her private experience is the only phenomenon accessible to an agent which she does not model with quantum mechanics and that the agent's awareness is the result of the experiment. Hence, there is no more ambiguity about when using the reduction postulate which does not say anything about the "real state" of the system that is measured but is nothing else than the updating of the agent's state assignment on the basis of her experience. In this respect, QBism is much clearer about what a measurement is than the relational interpretation which is dumb about that. In particular, there is no measurement when there is no agent: a Stern and Gerlach apparatus cannot measure by itself the spin of a particle. That is a very important point and we will see that Convivial Solipsism assumes something that is similar. QBists want to stress the fact that they depart from the standard Copenhagen interpretation in that a measurement is not "*an interaction between classical and quantum objects occurring apart from and independently of any observer*" [49]. They quote here the Landau and Lifshitz formulation of the Copenhagen interpretation. To be fair, this quotation is not really faithful to Bohr's position which is more subtle as we have seen. But it is true that even when it is taken in all its subtlety, Bohr's position is not able to solve the measurement problem in a satisfying way as we have seen since it does not make any explicit resort to the observer to define a measurement while simultaneously it appeals to the use that is done of the apparatus.

---

[21] I will not address here the part of QBism aiming at reconstructing the quantum formalism from symmetric informationally complete positive-operator-valued measure (SIC) which is not relevant for the present discussion.

[22] The subjective interpretation of probability has been mainly developed by de Finetti and Savage. It amounts to say that probabilities are related to an epistemic (hence personal and subjective) uncertainty and represent the degree of belief of an agent for the happening of an event.

For QBism, any agent can use the quantum formalism to model any system external to herself whether they be atoms, apparatuses or even other agents. For any agent, the personal internal awareness of other agents is inaccessible and not something she can apply quantum mechanics to. But the communication with other agents (verbal exchanges for example) is. That means that an agent can use a description of another agent through a superposed state encoding her probabilities for the possible answers to any question she might ask before getting a definite answer. This is very similar to the description we gave about the relation interpretation. A consequence is that it seems that reality can differ from an agent to another. But the same reservations as for the relational interpretation can be made about the meaning of the comparison between the perceptions of two agents since this comparison could only be done from a meta observer point of view, which does not exist. No third person point of view is allowed.

We will see that Convivial Solipsism shares many features with QBism, the most important one being the fundamental role that the observer plays in the measurement process and the relinquishment of any absolute description of the world. Now QBism is fuzzy on many aspects and leaves open many important questions whose answers would be necessary if one wants to get a detailed precise picture of what is going on. I give below a list of questions that naturally arise when one tries to go beyond a superficial understanding which at first sight seems satisfying but which raises many issues at second sight when one tries to understand more precisely what QBism means.

The first one is to clarify what it is that makes "something" an agent. According to QBists, an agent must have experience. Does that mean that an agent must be conscious? That is something that they are reluctant to accept. Nevertheless, they admit that experience and consciousness are difficult to disentangle[23]. But, they say that consciousness is not necessary for QBism and that experience is all is needed. However they admit that a computer programmed to use quantum mechanics would not be an agent and that the only agents known today are human beings… Acknowledging that an agent must be conscious to have experience while simultaneously refusing to discuss consciousness (which is not exactly the same concept than experience) on the basis that experience is enough and that the slight conceptual difference between consciousness and experience is irrelevant to quantum physics would be a perfectly acceptable position if it did not just so happen that the difference is relevant because it helps understanding some of the questions below.

The QBist solution to the measurement problem is that through her interaction with the external world, the agent has an experience which is the result of the measurement. There seems to remain no more ambiguity about what a measurement is since there is no measurement without agent. Once admitted

---

[23] Private communication.

that an agent creates a result through her experience, it becomes acceptable to claim that the role of quantum mechanics is only to give rules allowing the agent to update her old beliefs with the result she got and from that, to compute her new beliefs about future experience. But now the big question is: how is it possible for an agent to create a result? Where this result does come from? Is it something that can be described by the quantum formalism (independently of the fact that of course we all know what it is to have experience)? QBism seems to take the fact that the agent gets a unique (and classical) result as a basic given fact and does not make any attempt to give an explanation of it. Claiming that quantum mechanics is only a list of recipes for any agent to compute her beliefs for her future experience from her old beliefs updated by new results would be acceptable if precisely one of the main problem in quantum mechanics was not to explain how one unique result is obtained. Classical probability theory (whether Bayesian or not) can be viewed as a list of recipes to compute the probability of some well-defined events that are not explained themselves by probability theory. Probability theory is not expected to explain why a coin can land head or tail. The fact that a coin lands head or tail is taken as an external fact and when one such event happens, it allows updating the previous series of tosses if the user is willing to use it to modify her belief about the fact that the coin is biased or not. The situation for quantum mechanics is totally different since it is supposed to be a universal theory that must explain how a result is obtained during a measurement. But QBism gives no reason for that. Moreover no reference is made to the decoherence process for explaining that the result we get is conform to what we want to measure (preferred basis problem) and that we do not see macroscopic superpositions. This is just something that is accepted as such. QBism claims to have no need of decoherence [48]:

*"For QBism, it is not the emergence of classicality that needs to be explained, but the emergence of our new ways of manipulating, controlling, and interacting with matter that do."*

All that could be acceptable for a strongly instrumentalist position, claiming that the fact that a unique classical result is obtained is something that is just taken as a given fact (about which it has nothing to say) and that quantum mechanics is not a universal theory of the universe but is limited to the computation of beliefs about future experiments. In this case, QBism would loss a great part of its interest. But QBism puts at the center of its thesis the concept of experience of agents and it should be committed to answer all the questions that this concept raises. As far as classical subjective probabilities are concerned, the agent updates her assignment because she learns something new about the external world and what she learns is considered as real and independent of the agent. In QBism how does that happen?

*"A measurement does not, as the term unfortunately suggests, reveal a pre-existing state of affairs. It is an action on the world by an agent that results in the creation of an outcome."* [50].

Hence it seems that QBists endorse the idea that there is an external world on which each agent can act and that this action modifies the external world and creates a result which is her experience (her own private world). So QBists give the beginning of an explanation of how a result is obtained by an agent. Hence, they cannot refuse to go further in this explanation. The status of this external world is not clear. Is the external world one unique external world common to all the agents or has each agent her own external world in addition to her own personal world that she builds from her interactions with the external world?

If it is assumed that there is a unique external world as seems to indicate Fuchs when he says [48]:

*"… the real world, the one both are embedded in –with its objects and events – is taken for granted. What is not taken for granted is each agent's access to the part of it he has not touched"*

then a question arises about what happens in this external world when a measurement is made by an agent. The interaction between an agent and the external world has a special status since it is assumed to be out of the scope of quantum mechanics (that is the very reason why the measurement problem is solved). Is this interaction a physical interaction? If yes, is the external world changed? How? Why is the agent getting a unique result? The result participates to the construction of the agent's personal world. But is this result to be interpreted as real in the external world? If yes, it seems that we are back to the position of London and Bauer with a process that can have a physical impact on the world while resting outside of the scope of quantum mechanics. If on the contrary, the result is supposed to be only in the mind of the agent and the physical system is unaffected, then we arrive to something that is very similar to Convivial Solipsism except that, as we noticed, QBism is unable to explain why interferences between the different initial components of the state vector (which remains superposed) are not observed after the agent got a result while the system is physically unaffected. This is precisely what Convivial Solipsism explains.

Now, if each agent has her own external world, what is the difference between her external world and her personal world? Are the properties of a system seen by an agent real properties of the system inside the external world? It seems that it is not the case but what are they? Moreover, assume that Alice makes a measurement (say of the spin along Oz of an electron) and creates one result for her (say +) through her action on the world. Bob keeps a superposed state vector for the electron (and also for Alice who is entangled with the electron after the measurement) allowing him to compute the probability for getting + or – if he makes a measurement. But it is perfectly possible that Bob gets the result -. Even if we fall here in the trap that we denounced above of adopting a meta observer point of view in comparing the results got by Alice and Bob, it seems that each agent creates the values that she sees for each property of each system. The picture becomes that of separated worlds each one attached to one agent who acts on her own world to create what she sees. That could be not so far from Convivial Solipsism.

Even though, as we have seen, that would not be coherent to start mentioning these points and to refuse to take into consideration all the questions that these points raise, QBists could adopt an extreme instrumentalist position and answer that they are not interested in answering these questions. In this case, QBism would at least have the merit to claim that it is impossible to understand what science is (especially quantum mechanics but not only) without taking into account that ultimately, all what science describes is our personal experience. It also gives a clear prescription of how to use the recipes of quantum mechanics ascribing to use the reduction postulate only when an agent is involved. But this limited position is not really satisfying and QBism should aim at more.

At this stage, I do not pretend that there is any contradiction in the description given by QBism but only that all these points should be made clear in order to be able to get a global coherent picture of the ontology of QBism. All the above questions can find some kind of answer if consciousness is taken into consideration. The puzzling question about the creation of a result from the interaction between the agent and the external world (unacceptable if understood in a physical sense) and the nature of the construction of the personal world from the external world find natural explanations in Convivial Solipsism which is much more explicit on these points, as I will show in §8.

## 4. The EPR Paradox

It is well known that the initial Einstein, Podolski, Rosen paper [24] was aimed at proving that quantum mechanics is not complete. They first state the now famous Criterion of Reality:

*"If, without in any way disturbing a system, we can predict with certainty (i.e., with probability equal to unity) the value of a physical quantity, then there exists an element of reality corresponding to that quantity."*

Then they propose a thought experiment in which two quantum systems interact in such a way that two conservation laws hold, the relative position along the *x*-axis and the total linear momentum along that same axis which is always zero. They then say that measuring the position of the first system allows to predict with probability one the position of the second one. According to the Criterion of Reality, that means that the predicted position is an element of reality for the second system. The same reasoning is possible if this is the momentum of the first system that is measured. Hence, both position and momentum are elements of reality for the second system. This is in contradiction with quantum mechanics that claims, if it is complete, that there can be no simultaneously real values for incompatible quantities. To be correct, this demonstration needs nevertheless two additional postulates. The first one is separability: at the time when a measurement is made on the first system, the second system maintains its separate identity even though it is correlated with the first one. The second postulate is locality: at the time of measurement, the two systems no longer interact so no real change can take place in the second system as a consequence of a measurement made on the first one. As Fine puts it [51]:

*"In summary, the argument of EPR shows that if interacting systems satisfy separability and locality, then the description of systems provided by state vectors is not complete. This conclusion rests on a common interpretive principle, state vector reduction, and on the Criterion of Reality."*

Of course, the EPR argument was widely discussed at the time in particular by Bohr. In substance, Bohr acknowledges the fact that the measurement of the first system does not involve any mechanical disturbance of the second system. But he claims that this measurement on the first system does involve "an influence on the very conditions which define the possible types of predictions regarding the future behavior of the other system." Now measuring the position of the first system does not allow any prediction for the momentum of the second system. Since, as we have seen, for Bohr the very meaning of a property depends on the experimental set up that is used, it is not possible to speak of the momentum as being an element of reality of the second system if the position is what is measured on the first one. Hence for him, the argument does not hold anymore. Since it is no more possible to consider that both observables are defined simultaneously, position and momentum cannot have definite values at the same time and quantum mechanics can again be considered as complete. A careful analysis shows nevertheless that if Bohr succeeds in refuting the incompleteness of quantum mechanics this is at the price to giving up either separability or locality. Indeed, claiming that it becomes possible to speak of the position of the second system only as soon as the position of the first one has been measured, as far as the two systems can be from each other, means either that the two systems are actually one unique system before this measurement or that the first one can have a sort of instantaneous action at a distance on the second one.

It has become usual to state the EPR paradox not through the initial formulation with position and momentum as incompatible observables but through the formulation given by Bohm [52]. In Bohm's formulation, a spin zero particles decays with spin conservation into two spin ½ particles U and V in the singlet state of spin. Such a state can be written as:

$$|\psi\rangle = \tfrac{1}{\sqrt{2}}\left[\ |+\rangle^U\ |-\rangle^V\ -\ |-\rangle^U\ |+\rangle^V\ \right] \tag{12}$$

Where $|+\rangle$ and $|-\rangle$ are the state corresponding to a spin +1/2 and -1/2 of the related particles along an arbitrary chosen axis. The main point is that this state is invariant by rotation and that the total spin of the two particles along any axis must be zero. Hence, if a measurement of the spin of the particle U along one arbitrary axis is "+" then a measure of the spin of the particle V along the same axis will have to give "-". Now the same reasoning than the previous one with position and momentum in the initial formulation can be done. In this case, the incompatible observables are for example, the spin along x-axis and the spin along z-axis. Briefly stated, if a measurement of the spin along x-axis of U is made, this allows to predict with probability one the value of the spin along x-axis of V. Hence, according to EPR Criterion of Reality, the value of the spin along x-axis of V is an element of reality for V. As, this can be done also for the measurement of the spin along z-axis of U, that means that the value of the spin along z-axis of V is also an element of reality for V. Hence, both the value of the spin along x-axis and the value of the spin along z-axis of V are simultaneous elements of reality for V. But, this is in contradiction with the fact that quantum mechanics is complete since according to the quantum formalism, the observable spin along x-axis and the observable spin along z-axis are not commuting and cannot have simultaneously well-defined values.

Let's first make clear that the situation is not similar to the classical situation where "things are done" when the two particles separate. Assuming that, would mean that the explanation for the fact that the value of the spin along z-axis of V is "-" (resp. "+") if the measurement of the spin along z-axis of U have given the result "+" (resp. "-") is simply the fact that as soon as U and V separate, the value of the spin along z-axis of U was already "+" (resp. "-") and the value of the spin along z-axis of V was already "-" (resp. "+"). That would mean that the state of U and V right after their separation is either $|\psi_1\rangle = |+\rangle_z^U |-\rangle_z^V$ or $|\psi_2\rangle = |-\rangle_z^U |+\rangle_z^V$. Now, if we consider a set of N pairs U and V, that means that this set is a mixture of N/2 pairs in the state $|\psi_1\rangle$ and N/2 pairs in the state $|\psi_2\rangle$. But this is in contradiction with the fact that the pairs are in the singlet state. Indeed, the predictions given by N systems in the singlet state and those given by this mixture are different as soon as another axis is considered. For example, the probability for finding the same spin "+" along x-axis for U and V is ¼ in the case of the mixture while it is 0 in the case of the singlet state. This proves that it is not possible to consider that the value of the spin along any axis is already fixed as soon as U and V separate and before any measurement. It is only when a measurement is done on U along one particular axis that the value of the spin along this axis becomes defined.

Now comes the striking point: if this is true, that means that the value of the spin of V along the same axis is also only determined when the measurement on U is done, whatever the distance between U and V be. So a measurement of the spin of U along an axis R providing the result "+" has the effect of projecting the singlet state vector of the pair into a new state that is a tensorial product of two pure states corresponding to a definite value of the spin along this axis:

$$|\psi\rangle = \frac{1}{\sqrt{2}}\left[|+\rangle^U |-\rangle^V - |-\rangle^U |+\rangle^V\right] \xrightarrow{becomes} |+\rangle_R^U |-\rangle_R^V \qquad (13)$$

So, the state vector of V becomes $|-\rangle_R^V$ immediately after the measurement on U having given the result "+". Hence, this measurement has three effects. It determines the value of the spin of U along the axis R, it separates U and V allowing each of them to possess its own state (which was not the case before because they were entangled) and it determines the value of the spin of V along the axis R. This measurement has then an influence on the state of V and this influence at a distance is instantaneous. I will discuss in more details this point later but it is useful from now to notice that this conclusion is embarrassing especially for those realists who assume that the state vector is representing a real state of the system, for a change in the state vector is, for them, a change in the real state of the system. Hence it seems that locality is violated in some respects even if no mechanical disturbance is involved and even if it can be shown that is not possible to transmit any information through this process.

To be clear, let's summarize the reasoning at this stage: either quantum mechanics is not complete because things are already determined before the measurement (which is a situation that it is not possible to describe inside the quantum formalism and that leads to assuming hidden variables) or there is a violation of locality. This violation is more or less serious depending on the level one is assuming a realist position and whether one interprets the state vector as representing a real state of the system or not. The EPR argument was only a thought experiment until Bell came.

## 5. Bell's inequalities

Bell shows [53] that, under a given set of assumptions, certain of the correlations that can be measured between the two systems in an EPR experiment must obey some inequalities. The interesting point is that quantum theory predicts that Bell's inequalities must be violated. That means that quantum theory is inconsistent with the assumptions used to derive the inequalities.

We will follow a very simple proof given by Maccone [54] which is useful for understanding what Bell's inequalities mean without being involved into useless technical stuff. Let's call "local theory" a theory where the outcome of an experiment on a system are independent of the actions performed on a different system that has no causal connection with the first. Let's define as realist[24] a theory where it is meaningful to assign a property to a system independently of whether the measurement of this property is carried out. Then quantum mechanics is either non local or non-realist.

Let's consider three arbitrary two valued properties A, B, C and two systems satisfying locality and realism for these properties. That means that the measurement of one of these properties on one system has no effect on the measurement of anyone of these properties on the other system and that it is meaningful to consider that each system has definite values for each one of these properties even if no measurement is made. That is exactly what the case is in classical physics. Let's assume that the two systems are identical so that if the measurement of a property on the first one gives a certain result, it is sure that the measurement of the same property on the second one will yield the same result. Call $P_{same}(A,B)$ the probability that the property A of the first system and the property B of the second system have the same value (both 0 or both 1). Then for identical systems, $P_{same}(A,A) = P_{same}(B,B) = P_{same}(C,C) = 1$. Now it's easy to show that:

$$P_{same}(A,B) + P_{same}(A,C) + P_{same}(B,C) \geq 1 \qquad (14)$$

Maccone gives a graphical proof but that can be shown directly. We can rewrite the first member of the above equation as:

$P_{same}(A,B) + P_{same}(A,C) + P_{same}(B,C) = P(A=1, B=1, C=0) + P(A=1, B=1, C=1) + P(A=0, B=0, C=0) + P(A=0, B=0, C=1) + P(A=1, C=1, B=0) + P(A=1, C=1, B=1) + P(A=0, C=0, B=0) + P(A=0, C=0, B=1) + P(B=1, C=1, A=0) + P(B=1, C=1, A=1) + P(B=0, C=0, A=0) + P(B=0, C=0, A=1)$

$= 1 + P(A=1, B=1, C=1) + P(A=0, B=0, C=0) + P(A=0, C=0, B=0) + P(B=1, C=1, A=1) \geq 1$

$(15)$

Of course, this supposes that the probabilities are defined on a set of events that obeys the standard Kolmogorov axioms of probability, That means that the three properties have simultaneously definite values on each systems and that these values are what they are independently of any measurement of them (what realism and locality imply).

Now let's show that this inequality can be violated in some cases in quantum mechanics. Consider two spin ½ systems in an entangled state:

$$|\psi\rangle = \tfrac{1}{\sqrt{2}}\left[\, |+\rangle_z^1 |+\rangle_z^2 + |-\rangle_z^1 |-\rangle_z^2 \,\right] \qquad (16)$$

Let's consider the three observables A, B and C defined as: A is the spin along the z-axis, B is the spin along an axis u got from a rotation of π/3 of the z-axis in the Oxz plane, C is the spin along an axis v got from a rotation of -π/3 of the z-axis in the Oxz plane. Then for each system:

---

[24] Maccone is willing to avoid using the term "realist" because it is, according to him, too philosophically laden and he prefers using the term "counter-factual definite". Here I will continue using the term "realist" in the meaning of EPR criterion of reality.

$$|+\rangle_u = \frac{1}{2} |+\rangle_z + \frac{\sqrt{3}}{2} |-\rangle_z$$

$$|-\rangle_u = \frac{\sqrt{3}}{2} |+\rangle_z - \frac{1}{2} |-\rangle_z$$

(17)

$$|+\rangle_v = \frac{1}{2} |+\rangle_z - \frac{\sqrt{3}}{2} |-\rangle_z$$

$$|-\rangle_v = \frac{\sqrt{3}}{2} |+\rangle_z + \frac{1}{2} |-\rangle_z$$

It can be checked that:

$$|\psi\rangle = \frac{1}{\sqrt{2}}\left[|+\rangle_z^1 |+\rangle_z^2 + |-\rangle_z^1 |-\rangle_z^2\right] = \frac{1}{\sqrt{2}}\left[|+\rangle_u^1 |+\rangle_u^2 + |-\rangle_u^1 |-\rangle_u^2\right] = \frac{1}{\sqrt{2}}\left[|+\rangle_v^1 |+\rangle_v^2 + |-\rangle_v^1 |-\rangle_v^2\right] \quad (18)$$

Assuming realism and locality, which means that the two systems have the same values for all properties since if a measurement of one property on one system gives a certain result, any measurement of the same property on the other system will give the same result.

Now, computing for example $P_{same}(A,B)$ can be done through rewriting $|\psi\rangle$ as:

$$|\psi\rangle = \frac{1}{2\sqrt{2}}\left[|+\rangle_z^1 (|+\rangle_u^2 + \sqrt{3} |-\rangle_u^2) + |-\rangle_z^1 (\sqrt{3}|+\rangle_u^2 - |-\rangle_u^2)\right] \quad (19)$$

It is easy to see that according to the Born rule, the probability to get + for both properties is 1/8 and the same for the probability to get –. So, the probability to get the same value is ¼. Similar computations show that it is the same for $P_{same}(A,C)$ and $P_{same}(B,C)$. Hence:

$$P_{same}(A,B) + P_{same}(A,C) + P_{same}(B,C) = ¾ \leq 1 \quad (20)$$

in contradiction Bell's inequality. That means that quantum mechanics is incompatible with either realism or locality (or both).

The simple form of Bell's inequality we have presented here is not convenient to set up an experiment in order to test if the inequality is actually violated by quantum mechanics. Many different variant of Bell's inequality have been derived. The most usual form for the experiments is the Bell-Clauser-Horne-Shimony-Holt (BCHSH) inequality [55] and it concerns the polarization of photons. As is well known, even if it is always possible to question one or another subtle detail in each experiment to escape the conclusion, it is now widely agreed, after Aspect's experiments, that the inequalities are violated and that the result given by quantum mechanics is the correct one [56, 57].

What does that mean that no theory can respect both realism and locality? It is well known that the quantum mechanics formalism provides no way to describe the fact that incompatible observables (such as position and momentum or spin along two different axes) have simultaneously definite values. In this sense, quantum mechanics is not directly compatible with realism (at least understood according to the way presented above) and this is the reason why Einstein thought that it was not complete. But, it could be possible to imagine that it is possible to complete it with hidden variables which describe the "real" values possessed by the incompatible observables. That's what Bohm theory does as we saw §3.1.2. But of course, in order to be empirically correct, this theory must give the same predictions than ordinary quantum mechanics (and this is actually the case). So Bell's inequality is also violated in Bohm theory which means that locality is not respected (since realism seems to be[25]). More generally, any hidden variables theory must be non-local. That means that even if incompatible properties can have simultaneously definite values independently of any measurement, any measurement of one observable on one system can possibly affect instantaneously the value of the corresponding observable on another distant entangled system.

---

[25] As we saw in §3.1.2, the kind of realism that Bohm theory respects is nevertheless very peculiar.

If we summarize, EPR argument shows that either quantum mechanics is not complete or it is not local. Bell's inequality shows that any hidden variables theory must be non-local. Hence, non-locality seems unavoidable.

We will nevertheless see that Convivial Solipsism allows quantum mechanics to be local even if it is at the price of reinterpreting the formalism in a rather radical way.

## 6. Summary of the puzzling questions

At this stage, many questions are puzzling:

- What is a measurement and when must we use the Schrödinger equation or the reduction postulate for describing the evolution of systems?

- When and why only one of the many possible results of measuring an observable is selected?

- If the measurement does not reveal a preexisting value, how is it possible that this value be created during the measurement?

- Does this value so created belong to the system itself, does it belong to the system and the apparatus, does it concern the external reality and if so, is this reality the same for all observers, or is the value something that concerns merely the observer?

- If even macroscopic systems can become entangled, why don't we observe macroscopic superpositions?

- How do we know which observable is measured when we use an apparatus? (That is the preferred basis problem, see §7.3).

- How must we understand the non-locality shown by Bell's inequalities and is there any instantaneous action at a distance?

Some of the interpretations we have seen above give different answers to some of these questions. But none gives answers to all these questions in a coherent way. We are going to see that decoherence allows to answer the fifth and the sixth questions. Convivial Solipsism (which welcomes the decoherence mechanism) is an attempt to answer the others and to get a coherent general image even if the final picture it gives will perhaps appear a little weird to those attached to a more familiar view of reality.

## 7. The decoherence mechanism

I present here the decoherence mechanism that provides answers to some of the questions stated in §6. This solution came first from a remark from Zeh [58] that *no system is really totally isolated*. Hence it is necessary to take the environment into account. Then Zurek [59, 60] did the final move. The decoherence theory is nothing else than the description of the way to take the interaction between the system, the apparatus and the environment into account inside the quantum formalism. I shall first describe briefly the technical framework in which the decoherence theory is usually stated. Then I will explain how decoherence works and what it achieves (mainly selecting the preferred basis and explaining why macroscopic superpositions are never seen). I will refute then the realist interpretation of decoherence that pretends that decoherence solves the measurement problem.

### 7.1. The density matrix formalism

The density matrix formalism has been invented for being able to deal with statistical mixtures of systems being in different pure states, as in classical statistical mechanics. The density matrix (which is actually an operator) of a system being in a pure state $|\Psi_S\rangle$ is: $\varrho_S = |\Psi_S\rangle\langle\Psi_S|$. For the sake of simplicity, we will consider a two dimensional Hilbert space[26] with a basis $(|\varphi_1\rangle, |\varphi_2\rangle)$:

Let $|\Psi_S\rangle = \alpha |\varphi_1\rangle + \beta |\varphi_2\rangle$ then the density matrix in this basis is:

$$\varrho_S = \begin{pmatrix} |\alpha|^2 & \alpha\beta^* \\ \alpha^*\beta & |\beta|^2 \end{pmatrix} \tag{21}$$

In this case: $\varrho_S^2 = \varrho_S$ and $\text{Tr}(\varrho_S^2) = 1$. That means that for a pure state, the density operator is a projector.

If we consider now a statistical mixture E of N systems in the state $|\varphi_1\rangle$ with probability $p_1$ and in the state $|\varphi_2\rangle$ with probability $p_2 = 1-p_1$, the density matrix is:

$$\varrho_E = \begin{pmatrix} p_1 & 0 \\ 0 & p_2 \end{pmatrix} \tag{22}$$

This density matrix is diagonal so that its form is analogous to the classical case of a statistical mixture of a proportion $p_1$ of systems in the state $|\varphi_1\rangle$ and a proportion $p_2$ of systems in the state $|\varphi_2\rangle$.

In general, let's consider a statistical mixture of N systems in the states $|\Psi_k\rangle$ with probability $p_k$

$$|\Psi_k\rangle = \alpha_k |\varphi_1\rangle + \beta_k |\varphi_2\rangle \tag{23}$$

Then $\varrho_k = |\Psi_k\rangle\langle\Psi_k|$ and $\rho = \sum_k p_k \rho_k$. In this case $\varrho_S^2 \neq \varrho_S$. An important point is that it is always possible to find a basis ($|\zeta_1\rangle, |\zeta_2\rangle$) in which the density matrix is diagonal so that its form is analogous to the classical case of a statistical mixture of a proportion $\mu_1$ of systems in the state $|\zeta_1\rangle$ and a proportion $1-\mu_1$ of systems in the state $|\zeta_2\rangle$.

In general, the diagonal element $\rho_{ii}$ is the probability to find the system in the state $|\varphi_i\rangle$ and the non-diagonal element $\rho_{ij}$ is linked to the interferences between $|\varphi_i\rangle$ and $|\varphi_j\rangle$.

The mean value of any observable A is given by $\bar{A} = Tr(\rho A)$ and the frequency with which the measurement of A will give the eigenvalue $a_k$ is $w_k = Tr[\rho P(a_k)]$ where $P(a_k)$ is the projector onto the subspace spanned by the eigenvectors of A corresponding to the eigenvalue $a_k$.

The Schrödinger equation gives:

$$i\hbar \frac{d}{dt}\varrho(t) = [H(t), \varrho(t)] \tag{24}$$

Let's now consider for example, an ensemble of N identical systems S each composed of two entangled subsystem U and V and let the state

---

[26] Of course, all will be said here for a two-dimensional Hilber state is applicable for any other dimensional Hilbert space.

$$|\Psi_S\rangle = \sum c_{ij}|u_i\rangle v_j\rangle \tag{25}$$

be the state vector of each S. The density matrix describing this ensemble is $\varrho_S = |\Psi_S\rangle\langle\Psi_S|$. Assume now that we are interested only in the ensemble of the subsystems V. It is well known that in this case it is not possible to assign a state vector to V. Only the composed system S has a state vector. Now the quantum formalism allows to compute for this ensemble a density matrix from which it is possible to compute the mean value of any observable A and the frequency with which the measurement of A will give each one of its eigenvalues, through exactly the same formula as the one given above provided that A is acting only in the subspace linked to V. The ensemble V, though without state vector can be described by a density matrix. This density matrix is got by a mathematical operation called the partial trace of the global density matrix $\varrho_S$. This is a general procedure. When one is interested only in a subsystem of a composed global system, when one decides to restrict the measurements to observables that are related only to this subsystem, the density matrix that one must use is the partial trace of the global one. This will play a major role in the decoherence mechanism.

One very important conceptual point about this mathematical operation of partial trace has been shown by d'Espagnat [61]. Let's consider for example the density matrix of an ensemble E composed of N systems S of two spin ½ particles in the singlet state:

$$|\psi\rangle_E = \frac{1}{\sqrt{2}}\left[|+\rangle_z^U|-\rangle_z^V - |-\rangle_z^U|+\rangle_z^V\right] \tag{26}$$

and $\varrho_E = |\Psi_S\rangle\langle\Psi_S|$. The density matrix of the ensemble $E_U$ composed only of the U particles is got form the partial trace of $\varrho_E$:

$$\varrho_{E_U} = \frac{1}{\sqrt{2}}\left[|+\rangle_z^U\langle+|_z^U + |-\rangle_z^U\langle-|_z^U\right] \tag{27}$$

It has the same form as the density matrix of a mixture of particles U having a spin + along Oz for half of them and a spin – for the other half. But it would be wrong to consider that it is actually the density matrix of a proper mixture of that kind. Indeed, assuming so (and the same for the V particles) would mean that the ensemble E is composed of a mixture of particles U and V having each a well-defined value of spin along Oz. Actually it would be composed for one half, of pairs of one particle U having a spin + along Oz and one particle V having a spin – along Oz and for the other half of pairs of one particle U having a spin - along Oz and one particle V having a spin + along Oz (the other combinations are impossible in the singlet state). But such a mixture, let's call it E', has a density matrix $\varrho_E{'}$ that is different from $\varrho_E$ and that would lead to predictions relative to the correlations of measurement of spin along other directions than Oz that would be different from the predictions made from $\varrho_E$. For example a measurement of the spin of U and V along Ox, would get a probability 0 from $\varrho_E$ for finding both spin equal to + while this probability would be ¼ from $\varrho_E{'}$.

This shows that even if the density matrix of a subsystem obtained as the partial trace of the density matrix of a global system has the same form as the density matrix of a proper mixture of systems being each in one well defined eigenstate of those entering in the composition of the global entangled state, it is impossible to consider that this density matrix represent a proper mixture. D'Espagnat has called this type of mixture an improper mixture. This point will be important for analyzing what decoherence really achieves because that is precisely because of the illegitimate assimilation of improper mixtures to proper ones that some authors claimed that decoherence solved the measurement problem.

Another important point to notice is that no individual system can have a diagonal density matrix with more than one non null element. Indeed, an individual system is necessarily in a pure state. Now, as we have seen, if this state is $|\Psi_S\rangle = \alpha |\varphi_1\rangle + \beta |\varphi_2\rangle$ in the basis ($|\varphi_1\rangle$, $|\varphi_2\rangle$) then the density matrix in this basis is:

$$\varrho_S = \begin{pmatrix} |\alpha|^2 & \alpha\beta^* \\ \alpha^*\beta & |\beta|^2 \end{pmatrix} \quad (28)$$

Whereas if the state of the system is only one of the vector of the basis (for example $|\varphi_1\rangle$) then the density matrix is:

$$\varrho_S = \begin{pmatrix} 1 & 0 \\ 0 & 0 \end{pmatrix} \quad (29)$$

One can see that in neither case the density matrix can be:

$$\varrho_S = \begin{pmatrix} |\alpha|^2 & 0 \\ 0 & |\beta|^2 \end{pmatrix} \quad (30)$$

with $\alpha$ and $\beta$ both not null. Of course this is true for any space of higher dimensionality.

So, such a diagonal density matrix, when attached to an individual system, describes inevitably an improper mixture. This will be important in the following to show that the decoherence process applied to an individual system is not sufficient to explain the reduction of the state vector even if it leads to a diagonal density matrix.

### 7.2. The role of the environment

Following Zeh's remark, Zurek [59, 60] proposed the following mechanism to explain the reduction. Let's take the environment into account in the measurement process and consider a big system composed of the initial measured system plus the apparatus plus the environment.

$$\varrho_{SAE} = |\Psi_{SAE}\rangle\langle\Psi_{SAE}| \quad (31)$$

After the interaction, according to the Schrödinger equation:

$$\Psi_{SAE} = \sum c_i|\varphi_i\rangle|A_0\rangle|E_0\rangle \rightarrow \sum c_i|\varphi_i\rangle|A_i\rangle|E_i\rangle \quad (32)$$

As previously, we can assume a two dimensional space without loss of generality (and let $c_{1,2} = \alpha, \beta$). In the basis ($|\varphi_1\rangle|A_1\rangle|E_1\rangle$, $|\varphi_2\rangle|A_2\rangle|E_2\rangle$) we have, similarly to equation (28):

$$\varrho_{SAE} = \begin{pmatrix} |\alpha|^2 & \alpha\beta^* \\ \alpha^*\beta & |\beta|^2 \end{pmatrix} \quad (33)$$

Apparently nothing has been gained! In the basis of the Hilbert space which is the tensorial product of the Hilbert space of the system plus the apparatus plus the environment, the density matrix has exactly the same form as before.

But the key point comes from the remark that we cannot perform measurements on all the degrees of freedom of the environment because that would require apparatuses that are totally out of reach. As we have seen, the quantum formalism prescribes in this case that the density matrix of the sub system SA formed by the initial system and the apparatus is given by the partial trace on the degrees of freedom of the environment of $\varrho_{SAE}$ which can be computed as:

$$Tr_E \varrho_{SAE} = \varrho_{SA} = \begin{pmatrix} |\alpha|^2 & Z\alpha\beta^* \\ Z\alpha^*\beta & |\beta|^2 \end{pmatrix} \tag{34}$$

Now, it is possible to show that in general the coefficient $Z(t)$ decreases towards 0 very rapidly. So:

$$\varrho_{SA}(t) = \begin{pmatrix} |\alpha|^2 & Z(t)\alpha\beta^* \\ Z(t)\alpha^*\beta & |\beta|^2 \end{pmatrix} \rightarrow \begin{pmatrix} |\alpha|^2 & 0 \\ 0 & |\beta|^2 \end{pmatrix} \tag{35}$$

This density matrix looks like the density matrix of the equation (22) that describes a statistical mixture and no more a pure superposed state. So it seems that each system belonging to the set of systems described by $\varrho_{SA}(t)$ has now a definite state corresponding to one of the eigenvectors of the observable that has been measured. This is the reason why many authors (including Zurek in his first papers) thought that the decoherence process allows to explain in an objective way the reduction of the state vector. We analyze in §7.4 the reasons why this is not correct.

### 7.3. The preferred basis

As we have seen previously, a measurement is described as an interaction between a system and an apparatus. Let the system S be in a state $|\Psi_S\rangle = \sum c_i |\varphi_i\rangle$ and the apparatus A be in the initial state $|A_0\rangle$. Then, before they interact, the state of the system – apparatus is the tensorial product:

$$|\Psi_{SA}\rangle = |\Psi_S\rangle \otimes |A_0\rangle = \sum c_i |\varphi_i\rangle \otimes |A_0\rangle \tag{36}$$

The usual description assumes that the apparatus is built in such a way that if the measurement is made on a system that is in the state $|\varphi_i\rangle$, the apparatus will be in the state $|A_i\rangle$ after the measurement whatever its initial state. The Schrodinger equation gives:

$$|\Psi_{SA}\rangle = \sum c_i |\varphi_i\rangle \otimes |A_0\rangle \rightarrow \sum c_i |\varphi_i\rangle \otimes |A_i\rangle \tag{37}$$

The reduction postulate says that if the outcome seen by the observer is one of the $|A_i\rangle$ then the state of the system after the measurement is the corresponding $|\varphi_i\rangle$ and the value of the observable that is measured is the eigenvalue corresponding to that $|\varphi_i\rangle$. What is implicit in this story is that all the process is described in the basis $|\varphi_i\rangle|A_i\rangle$ where the $|\varphi_i\rangle$ are supposed to be the eingenvectors of the observable that is measured and the $|A_i\rangle$ are supposed to be well defined macroscopic states of the apparatus. But we know that there are infinitely many basis of a Hilbert space. The same process could be described in any other basis which involves superposed states of the apparatus. Nothing in the quantum rules imposes that the basis that is used corresponds to the well-defined macroscopic states of the apparatus (the preferred basis). This is only because we know what we can observe that we choose this basis. Let's take the example of a spin ½ particle and a Stern and Gerlach apparatus whose magnetic field is oriented along Oz and hence measuring the spin along Oz. The preferred basis is: $\{ |+\rangle_z |\uparrow\rangle , |-\rangle_z |\downarrow\rangle \}$ where $|+\rangle_z$ (resp. $|-\rangle_z$) is the eigenvector of the observable spin along Oz with the eigenvalue +1/2 (resp. - 1/2) and $|\uparrow\rangle$ (resp. $|\downarrow\rangle$) is the state of the apparatus in which the particle has an impact at the top (resp. at the bottom) of the screen. In this basis, the measurement process is described as follows:

The initial state of the particle is $|\Psi_S\rangle = c_+ |+\rangle_z + c_- |-\rangle_z$. The Schrodinger equation gives:

$$|\Psi_{SA}\rangle = (c_+ |+\rangle_z + c_- |-\rangle_z)|A_0\rangle \rightarrow c_+ |+\rangle_z |\uparrow\rangle + c_- |-\rangle_z |\downarrow\rangle \tag{38}$$

So, the previous analysis applies and we have a measurement of spin +1/2 along Ox if we observe an impact at the top of the screen, described by the state $|\uparrow\rangle$ and -1/2 if we observe an impact at the bottom of the screen, described by the state $|\downarrow\rangle$.

But assume for example that $c_+ = -c_- = 1/\sqrt{2}$. Then $|\Psi_S\rangle$ is invariant under rotations and can also be written in the basis using the eigenstates of the spin along Ox:

$$|\Psi_S\rangle = 1/\sqrt{2}|+\rangle_z - 1/\sqrt{2}|-\rangle_z = 1/\sqrt{2}|+\rangle_x - 1/\sqrt{2}|-\rangle_x \tag{39}$$

In this case, $|\Psi_{SA}\rangle$ becomes:

$$|\Psi_{SA}\rangle = 1/\sqrt{2}|+\rangle_x |\Uparrow\rangle - 1/\sqrt{2}|-\rangle_x |\Downarrow\rangle \tag{40}$$

Where $|\Uparrow\rangle = 1/\sqrt{2}|\uparrow\rangle + 1/\sqrt{2}|\downarrow\rangle$ and $|\Downarrow\rangle = 1/\sqrt{2}|\uparrow\rangle - 1/\sqrt{2}|\downarrow\rangle$

The states $|\Uparrow\rangle$ and $|\Downarrow\rangle$ of the apparatus correspond to superposed impacts on the screen. Of course such states have never been observed but nothing in the formalism says that they must be eliminated. This way of writing the measurement is also legitimate and can be interpreted as a measurement of spin +1/2 along Ox if we observe a superposed impact $|\Uparrow\rangle$ and -1/2 if we observe a superposed impact $|\Downarrow\rangle$. What is the reason why the correct description is the one done with the preferred basis? In other terms, why is it impossible to measure the spin along Ox with a Stern and Gerlach apparatus whose magnetic field is oriented along Oz? The quantum formalism has nothing to say about this question.

The decoherence theory explains that the reduced density matrix ends up being diagonal in the eigenspace of an observable of the apparatus that commutes with the Hamiltonian of interaction between the apparatus and the environment. The observable pointer position is such an observable and the commutation means that it will be a constant of motion of the interaction Hamiltonian so that the interaction with environment will leave it unperturbed. Decoherence hence provides a natural explanation for the selection of the preferred basis.

Let's notice that decoherence does not apply to a system composed of two electrons where one is the system and the other would be assumed to play the role of an apparatus, as is supposed in the relational interpretation. So, as we noticed in §3.4, it seems difficult to assume that a measurement has been done when two electrons having interacted are in the singlet state. What is the direction along which the spin of the first electron is supposed to have been measured relative to the second one? This casts some doubt about the validity of this interpretation. The same objection has often been made against Everett's interpretation about the ambiguity of what constitutes the branches.

### 7.4. The realist interpretation of decoherence

This is the position of those [62, 63] who want to see decoherence as solving the measurement problem. It consists in considering that:

- first, the transition of the density operator given in equation (35) is justified and hence that the diagonal form

$$\begin{pmatrix} |\alpha|^2 & 0 \\ 0 & |\beta|^2 \end{pmatrix} \tag{41}$$

Is the correct form to use for predictions.

- second, this diagonal form is to be interpreted as a proper mixture of systems each of them being in a definite state (the states of the basis in which the density matrix is diagonal). Hence, the squares of the coefficients α and β are exactly like classical probabilities and are linked to the proportion of systems being in one state or the other.

This interpretation relies on several wrong assumptions and is not acceptable at all [7, 26, 64, 65, 66]. First of all, the final diagonal form of the density matrix is the result of the partial trace of the global density matrix and the reason why this partial trace can be done is entirely due to the fact that it is acknowledged that no measurement of the environment is possible for the observer. It gives the correct predictions provided that the observer will not do any measurement on the environment. That means that the final diagonal form of the density matrix is the form it takes for an observer with limited means of measurement. Hence, it is not an objective (without any mention of observer) reduction. Second, the small non diagonal terms that have been considered as null ($Z(t) \rightarrow 0$) are actually not rigorously null and can even become again big after a (very) long time. Considering that the diagonal form corresponds to the real systems means that the non-diagonal elements which are small but not null, are arbitrarily neglected. Those doing that argue that the effects coming from these elements are impossible to observe in practice and hence that for all practical purposes (FAPP) the diagonal matrix is equivalent to the non diagonal one. This is true FAPP but false in principle. That is not because something cannot be observed that it does not exist. So the replacement of the non-diagonal matrix by the diagonal one is illegitimate (not for computing but for inferring from it any conceptual conclusion).

Now, even if we consider the diagonal matrix, it is illegitimate to consider that it is associated with a proper mixture of systems each of them being in a definite state. This matrix has been obtained by taking the partial trace of the global operator and in this case, even if its form is similar to the form of a density operator of a proper mixture, that is not the case. As we have seen above, it is associated with an improper mixture. An improper mixture is composed of systems that are all identical. As Bell emphasized [3], the correct interpretation should be that each system is in a state where all the possibilities are simultaneously present. This is the celebrated "and / or" difficulty. So it is not correct to interpret the decoherence process as leading to a set of systems each having a definite state in a proportion given by the diagonal coefficients of the matrix.

This is even clearer if we consider the case of an individual system. We noticed that no individual system can have a diagonal density matrix with more than one element not null. So the diagonal density matrix got after the partial trace for an individual system must be understood as a mere tool for computing probabilities of results in case of a measurement but it is illegitimate to use it to infer something on the real state of the system.

Moreover, decoherence cannot allow to get rid of the reduction postulate as it is clear from the consideration of a repeated measurement on the same individual system. For decoherence, without this postulate, is dumb about the density matrix that is to be used after having got a definite result. So, repeating the same measurement without having projected the state of the system on the eigenvector corresponding to the result that has been got, would not guarantee that the same result will occur. In any case, it is difficult to see how it could be possible to eliminate the reduction postulate without anything replacing it since the preparation of systems (which lies at the very heart of quantum mechanics) relies on the fact that in order to assign a state vector to a system, it is first necessary to have made some measurements on the system and observed which value are got.

For all these reasons, the realist interpretation of decoherence is not acceptable and must be replaced by a more modest interpretation that I explain below.

### 7.5. What decoherence brings and doesn't bring

Solving the measurement problem would mean that, independently of any observer, the initially superposed state of the system has been reduced to a definite state. We have seen that it is not the case.

What is justified to say is that the diagonal form of the density operator is the one that can be used to compute (through the usual rules of quantum mechanics including the reduction postulate that cannot be forgotten) the probabilities of each result in case of a measurement. Indeed:

a) The partial trace done on the global operator is justified according to the rule explaining how to deal with the density operator when one does not observe some degrees of freedom or when one limits oneself to observe a subsystem of a composed system.

b) Neglecting the very small non-diagonal elements is justified by the fact that if we compute the probabilities of interferences due to them, we can see that they are negligible in practice and will never be observed. Indeed, we can actually keep the non-diagonal terms and compute the probability to observe these macroscopic interference effects. The computation of these probabilities shows that we will never observe them in practice. Hence keeping the diagonal form of the density matrix is equivalent FAPP to using the fully non-diagonal form.

So, decoherence explains why macroscopic interferences are never observed. It brings an explanation of the classical appearance of the world, provided we use the standard recipes to compute. It explains why we (human observers) cannot observe any macroscopic superposition and why what we see is conform to the classical description of the world. But the underlying reality (if there is any) remains in a superposed and entangled state. If this is taken literally, that means that the reduction postulate is nothing but a convenient and practical way to describe the observations but does not correspond to any real physical process.

Now, the standard recipes to compute assume that we know what a measurement is. It is when a measurement is made that the probability of finding (observing) a specific result is given by the corresponding diagonal element of the density matrix. But nowhere inside the formalism of decoherence it is said what a measurement is. Decoherence has nothing to say about the reason why a superposed state gives only one result among the many possible ones when a measurement is done. Hence the postulate according to which only eigenvalues can be obtained during a measurement has to be conserved and the reduction postulate is needed for explaining why a unique value is observed and for insuring that immediately repeating a measurement will provide the same result. Hence the problem of knowing what a measurement is and when it occurs remains a mystery. In particular, the hope to get rid of any mention of observer in the formulation of quantum mechanics is still not fulfilled at all.

### 8. Convivial Solipsism

Even though today a large number of physicists would still like to consider that science is strongly objective and would not even be willing mentioning consciousness, many great physicists of the past insisted on the important role that consciousness plays. Planck said [67]:

*"I regard consciousness as fundamental. I regard matter as derivative from consciousness. We cannot get behind consciousness. Everything that we talk about, everything that we regard as existing, postulates consciousness."*[27]

---

[27] I am indebted to Chris Fuchs for this quotation from Planck.

The conclusion of all what precedes is that it seems impossible[28] to develop a coherent interpretation of the quantum formalism that avoids making any reference to the concept of an observer endowed with a consciousness.

Actually it is easy to see that the measurement problem arises in a realist context where one thinks that the state vector is representing the physical state of the system independently of any observer and hence, such that it is the same for all the observers. It is then considered as an absolute entity. It is in this framework that the measurement problem has been formulated in §2.1 following von Neumann. The astonishment that the two postulates of evolution give different results is natural as soon as it is considered that the two ways of describing the measurement are possible and that the two different final state vectors they provide concern the physical state of the system that should be the same for all the observers. But if the state vector is now considered to be relative to each observer and if it is clear that the reduction postulate must be used only when this very observer becomes aware of one result, then there is no more problem.

As Bitbol says [68]:

*"The measurement problem boils down to finding a way to articulate the indefinite chain of relational statements of the quantum theory to the absolute statements that are used in the experimental work. An articulation of this kind can easily be found, provided one realizes that the latter absolute statements are in fact indexical; provided one realizes that these statements are only absolute relative to us [...]. At this point one is bound to realize the ineliminability of situatedness from the apparent neutral descriptions of quantum mechanics."*

Convivial Solipsism[29] draws all the consequences of these ideas and rests on two main assumptions completed by the use of decoherence.

### 8.1. The first assumption: the hanging up mechanism

Acknowledging the seeming impossibility to get rid of the concept of consciousness in the measurement problem, the first assumption states that a measurement is but the awareness of a result by a conscious observer. In Convivial Solipsism, a measurement occurs when (and only when) a conscious observer becomes aware of a result. This cancels the ambiguity about when to use the Schrödinger equation and when to use the reduction postulate. That looks like the old proposal of London and Bauer and Wigner. But, if we take the decoherence process into account, there is now a big difference with their position. They thought that the reduction that occurs during a measurement was a physical process through a real action of the mind on the system, the mind changing the real physical state of the system. Convivial Solipsism defends a much less shocking position. Let's remind that at the end of the process of decoherence the reduced density matrix is diagonal in the preferred basis. We insisted on the fact that it does not mean that the measurement problem is solved since this diagonal density matrix is not the density matrix of a proper mixture of systems each one in one state of the preferred basis. The systems remain in a superposed and entangled state and the reduced density matrix is only a tool giving the probabilities for observations. No other physical meaning should be attributed to the reduced density matrix. Now, the diagonal form shows that no interference between the states of the preferred basis is observable. It remains to explain why only one of the different possible outcomes is observed (the famous Bell's problem "and / or") and this is precisely what the hanging up mechanism does through the selection of one result as I explain below. The reduction is then a way to describe what appears to the observer and does not affect the "reality" (whatever the meaning of this term that we will investigate further) which remains superposed.

---

[28] I recall that this conclusion is valid inside the context of the standard quantum mechanics assuming that we don't consider the modifications mentioned in §3.1.

[29] Convivial Solipsism is a widely modified and extended version of a first model initially proposed by d'Espagnat [7, 61]. I have first stated it in a 2000 book [69]. In a way, it shares some features with the initial position of Everett (but it does not agree with the many-worlds interpretation given by Graham and De Witt) except that it clarifies some points that were unclear in Everett's formulation and that, contrary to Everett's description, it is a one world theory and it is not susceptible to be given a simple realist meaning.

The reduction is not a physical process but merely the fact that when a conscious mind makes an observation, what it can see is only one of the results described by the diagonal density matrix. That does not mean that the state of the system is physically reduced (actually the system remains in an entangled state with the apparatus and the environment) but that the conscious mind can only be aware of one result that is selected at random following the Born rule.

Let's recall what happens in the initial Everett's interpretation (not in the "many-worlds" interpretation that I consider to be undermined by decoherence since it eliminates the possibility to observe, even in principle, interferences between the different branches which are causally disconnected). There is no reduction (the physical world remains in a superposed state) but the memory of the observer is different according to the branches corresponding to possible results.

If $|O_0\rangle$ is the initial state of the observer and $|O_i\rangle$ the i$^{th}$ state of the observer's memory:

$$\Psi_{SAEO} = \sum c_i |\varphi_i\rangle |A_0\rangle |E_0\rangle |O_0\rangle \rightarrow \sum c_i |\varphi_i\rangle |A_i\rangle |E_i\rangle |O_i\rangle \tag{42}$$

For Everett, each $|O_i\rangle$ corresponds to an experience of the observer having the feeling (or the memory) to have observed the result $|A_i\rangle$.

Now, contrarily to what happens in Everett's interpretation where the observer has as many relative states in which he is aware of one definite result as there are possible results, in Convivial Solipsism, we assume that the observer is aware of only one result which is selected at random. I call that the "hanging-up mechanism". Of course, the hanging-up mechanism has exactly the same status (it must be postulated and one cannot do without it) and a role similar (but slightly weaker as we will see in §8.3) than the reduction postulate. It is a weak reformulation of it. But the conditions of its usage are clearly given and there is no more ambiguity about when to use it.

So, the first assumption is:

"A measurement is the awareness of a result by a conscious observer whose consciousness selects at random (according to the Born rule) one branch of the entangled state vector written in the preferred basis and hangs-up to it. Once the consciousness is hung-up to one branch, it will hang-up only to branches that are daughters of this branch for all the following observations."

The last part of the first assumption guarantees that repeating the same measurement will give again the same result.

Assume for example that the system is a spin 1/2 particle in a superposed state along Oz.

$$|\Psi_S\rangle = \alpha |+\rangle_z + \beta |-\rangle_z \tag{43}$$

After the interaction with the apparatus and taking the environment into account the global state is:

$$|\Psi_{SAE}\rangle = \alpha |+\rangle_z |\uparrow\rangle |E_+\rangle + \beta |-\rangle_z |\downarrow\rangle |E_-\rangle \tag{44}$$

If we include the state of the observer we get:

$$|\Psi_{SAEO}\rangle = \alpha |+\rangle_z |\uparrow\rangle |E_+\rangle |☺\rangle + \beta |-\rangle_z |\downarrow\rangle |E_-\rangle |☹\rangle \tag{45}$$

We have here to make a difference between the physical brain of the observer and her consciousness. $|☺\rangle$ and $|☹\rangle$ are the state of the brain of the observer. Now, the hanging up mechanism says that the

consciousness of the observer chooses one branch at random (respecting the Born rule linked to the coefficients of the linear combination). Let's denote by $\widetilde{☺}$ the fact to be aware of having seen "+" and by $\widetilde{☹}$ the fact to be aware of having seen "-" then after the hanging up mechanism, either $\widetilde{☺}$ or $\widetilde{☹}$.

Here we must be very clear about not to confuse $|☺\rangle$ with $\widetilde{☺}$ and $|☹\rangle$ with $\widetilde{☹}$. $|☺\rangle$ and $|☹\rangle$ are physical states of the observer's brain that enter into the entangled universal sate vector. $\widetilde{☺}$ and $\widetilde{☹}$ are not state vectors and cannot enter into any linear combination with state vectors. That is the reason why they are not written as kets. They are just representing non physical states of awareness.

So, even if the universal entangled wave function is not reduced and remains as written in equation (45), for all subsequent measurements, everything happens for this observer as if the wave function was reduced either to $|+\rangle_z |↑\rangle |E_+\rangle |☺\rangle$ if her state of awareness is $\widetilde{☺}$ or to $|-\rangle_z |↓\rangle |E_-\rangle |☹\rangle$ if her state of awareness is $\widetilde{☹}$. This insures that repeating the same measurement will give the same result.

Assume that $\widetilde{☺}$. What happens if we measure now an observable which does not commute with the spin along Oz, for example the spin along Ox? Let's denote now with index 1 what is related to the results of the first measurement along Oz and with index 2 what is related to the second measurement along Ox. So we have $\widetilde{☺}_1$ after the first measurement. After the second measurement, the final entangled state will be:

$$|\Psi_{SAEO}\rangle = \frac{\alpha}{\sqrt{2}} |☺\rangle_1 \left[ |+\rangle_x |↑\rangle_1 |E_+\rangle_1 |↑\rangle_2 |E_+\rangle_2 |☺\rangle_2 + |-\rangle_x |↑\rangle_1 |E_+\rangle_1 |↓\rangle_2 |E_-\rangle_2 |☹\rangle_2 \right] + $$
$$\frac{\beta}{\sqrt{2}} |☹\rangle_1 \left[ |+\rangle_x |↓\rangle_1 |E_-\rangle_1 |↑\rangle_2 |E_+\rangle_2 |☺\rangle_2 - |-\rangle_x |↓\rangle_1 |E_-\rangle_1 |↓\rangle_2 |E_-\rangle_2 |☹\rangle_2 \right] \quad (46)$$

The hanging-up mechanism says that since we have $\widetilde{☺}_1$ after the first measurement, the selection for the second measurement must be done among the states that are correlated to $|☺\rangle_1$. The only term to consider is then:

$$|+\rangle_x |↑\rangle_1 |E_+\rangle_1 |↑\rangle_2 |E_+\rangle_2 |☺\rangle_1 |☺\rangle_2 + |-\rangle_x |↑\rangle_1 |E_+\rangle_1 |↓\rangle_2 |E_-\rangle_2 |☺\rangle_1 |☹\rangle_2 \quad (47)$$

So the final state of awareness will be either $\widetilde{☺}_1\widetilde{☺}_2$ corresponding to a spin + along Ox or $\widetilde{☺}_1\widetilde{☹}_2$ corresponding to a spin – along Ox (each one with a probability ½) which is conform to the usual prediction stating that if one performs a measurement of spin along Ox on a particle in a state + of spin along Oz, the possible results are + and – with probability ½.

Nevertheless, the physical state of the observer's brain continues to be superposed and entangled as described in equation (46) with the rest of the universe) even if the observer cannot be conscious of what happens in the other branches.

Of course that is totally unfaithful to the spirit of Everett who wanted to eliminate the reduction postulate. But it is greatly more economic regarding the number of conscious states of one observer and as soon as we are able to give first a rule stating when it must be used and second a reason why this hanging-up mechanism occurs, the repugnant aspect[30] of the reduction postulate becomes more acceptable. So let's give a simplified image. A pictorial way to understand that would be to think that among the many possibilities of the superposed state of the brain (however chosen only among those belonging to the preferred basis), only one of them is accessible to the consciousness very similarly to what could happen if, watching to a multi colored picture through one pair of colored glasses, the observer would see the picture colored with the unique same color than the glasses. Moreover, imagine that the observer draws one colored pair of glasses at random from an urn full of pairs of glasses of different colors, the color she would think the picture is would be predictable only in a probabilistic way. The physical brain of the observer is in an entangled state but the consciousness of the observer, due to internal filters that play the same role than the colored glasses, is able to hang-up to only one of the components of the entangled state. The filter that is selected is chosen at random following the probabilistic law given by the coefficient $c_i$ and so the Born rule is respected[31]. So, even if the wave function is never reduced, as in Everett interpretation, a great advantage of Convivial Solipsism is that it is not subject to the same problem for recovering the Born rule.

---

[30] According to some letters, Everett considered Bohr's approach as "somewhat repugnant" [70].

The question is then whether there can be any conflict between different observers hung-up to different branches. The answer is no for, as d'Espagnat puts it:

"A*ny transfer of information from B to A – for example, any answer made by B to a question asked by A – unavoidably proceeds through physical means. Therefore it necessary takes the form of a measurement made by A on B. And we know that under these conditions A necessarily gets a response (answer) that agrees with his own perception*" [61].

For any observer, everything outside her own private consciousness has to be treated as a quantum system obeying quantum mechanics. This is true of course of electrons but also of macroscopic objects and even of other conscious observers. This is quite similar to what we described for the relational interpretation and QBism. Hence, when an observer, say Alice, speaks with another one, Bob, it is as if Alice was doing a measurement on Bob. Let's take an example: the measurement of the spin along Oz of an electron in an initial superposed state of spin. Suppose Bob has performed such a measurement on this electron. From Bob's point of view, a measurement has been made and he knows the value of the spin along Oz of this electron. According to the hanging-up mechanism stated above, Bob's consciousness is hung-up to one of the two possible branches linked with the results up or down. From Alice's point of view, Bob becomes entangled with the electron, as described by the Schrödinger equation. That's again Wigner's friend problem. Now Alice can perform the same measurement on the electron. And Alice's consciousness will be hung-up as well to one of the two branches and will see one value. The crucial point is that this branch includes the state of Bob that is linked to the same value. So when Alice, hung-up to that branch, speaks with Bob to know what Bob saw, she performs a measurement on Bob and cannot hear Bob saying anything else than the same value that she got before. That's exactly the same mechanism than repeating the same measurement on a system initially in a superposed state. The second measurement will inevitably yield the same result. Alice will never hear Bob saying that he saw up when she saw down. No conflict is possible.

Now, the unavoidable question comes: Is it possible that Bob saw "up" and Alice "down" even if Alice will never know? Actually, as we noticed before, this question assumes that we adopt a meta point of view allowing us to speak simultaneously of what Bob and Alice saw. This meta point of view (third person point of view) is like God view, an absolute view assuming that it is meaningful to speak as a meta observer using absolute independent states. But in Convivial Solipsism, such an absolute meta point of view does not exist. So, even if it is tempting to adopt this meta level, to wonder if Alice and Bob can "really" see different things, has no meaning. Juxtaposing different points of view is not allowed. Hence, the intersubjective agreement is preserved because the communication between different observers is nothing but a measurement of one observer on the other. But this aspect, that we already discussed in §3.4.2 and §3.4.3 about the relational interpretation and QBism, is much clearer if we go right to the end of what that means. This is the second assumption.

---

[31] Actually this description could let think that the observer is facing the world which is given in front of her eyes. That is absolutely not what is assumed in Convivial Solipsism as we will see in §8.2. The explanation given here must be taken as a very limited analogy to help understanding the filters which prevent us to be in superposed states of consciousness. Indeed, the hanging-up mechanism has to be seen as a co-construction, from the world and the mind, of the result as we explain in §8.2.

### 8.2. The second assumption

All what has been said until now, could be understood as if there was a unique independent real world that all the observers witness (even if they are part of it) and which is described by the entangled global wave function that is never reduced. Of course, each observer would have her own private conscious experience of this universe (that could be different from one observer to the other depending on the branch to which she is hung-up) but the global wave function would be the same for all the observers and would be "The" wave function of a universe conceived as unique and independent of the observers. That is the point of view of Everett's interpretation[32]. In this case, we would recover a certain kind of realism even if the universe so described would be very different from the universe that each observer perceives. But Convivial Solipsism shares with QBism and the relational interpretation the idea that the entities inside the quantum formalism (state vectors and probabilities) are relative.

The second assumption is then:

"Any state vector is relative to a given conscious observer and cannot be considered in an absolute way."

That is similar to the assumption that state vectors are relative to systems (relational interpretation) or to agents (QBism). But the difference is that Convivial Solipsism assumes that a conscious observer is necessary.

In Convivial Solipsism, the state vectors, including the universal entangled one, are relative to one conscious observer. That means that even the universal entangled wave function has no absolute and universal validity but is relative to each observer. It represents only the description of the universe for a specific observer. In this case, one may ask what that means to continue speaking of a universe. The universe is no more an absolute reality existing outside and independently of any observer but is relative to each observer. That does not mean however that nothing else than the mind exists and that the universe is totally created by the consciousness of each observer. That would amount to coming back to a pure idealist position which is not what Convivial Solipsism lauds. Indeed, it is well known that pure idealism faces many difficulties.

Convivial Solipsism rather assumes that "there is something" else than consciousness, something that is not appropriate to talk of (similar to the Kantian thing in itself) and that consciousness[33] and "this something" give rise to what each observer thinks it is her reality, following Putnam's famous statement "the mind and the world jointly make up the mind and the world". So perception is not a passive affair: perceiving is not simply witnessing what is in front of us but is creating what we perceive through a co-construction from the world and the mind. In this respect, Convivial Solipsism is close to QBism since it considers that an observation does not reveal a preexisting state of affair but is a creation. But where QBism is dumb about how this creation can happen, Convivial Solipsism explains that each conscious observer builds her own world through the hanging-up mechanism[34].

---

[32] Actually, this is more conform to the initial point of view of Everett and not to the many-worlds interpretation given by Graham and De Witt.

[33] I will let here the concept of consciousness unanalyzed and take it as a basic given fact. It is enough to know what I mean when I say that I am conscious. I know and we all know what that means to perceive something in a conscious way. At the price of risking to be accused of instrumentalism, I would say that nothing more is needed here. I just want to stress the point that this does not mean that I support any version of dualism. But going further would lead to too speculative and unnecessary assumptions out of the scope of the present paper.

[34] Actually the ontology of Convivial Solipsism and the joint process of construction of the mind and the world are slightly more sophisticated and will be described in more details in a forthcoming more philosophically inclined paper. The simplified version given here is however faithful enough in the context of this paper to not distort the fundamental underlying idea.

That is a sort of solipsism because the consciousness of each observer is located inside its own branch of its own relative universal state vector independently of the others. But that is not a true solipsism as it welcomes both others minds and an external stuff that is independent of the mind. It is closer to Kant transcendental idealism. Now, it is convivial since no conflict is possible: the quantum rules together with the hanging-up mechanism for each observer prevent any possibility to notice a divergence between the perceptions of two different observers.

There is another striking consequence which is a strange answer to the famous phrase of Einstein: God does not play dice. Einstein was right, God does not play dice but each of us does! This is so because the random aspect of the quantum predictions comes, not from the fact that the physical systems change at random (the dynamic of the Universe, even if relative, is fully deterministic) but from the random way your consciousness chooses the branch to which it hangs-up!

### 8.3. The hanging-up mechanism reformulated and the contextuality of quantum mechanics

Now there is a subtlety that I have passed over in the way I presented the hanging-up mechanism above. The correct way to understand the hanging-up mechanism is a little bit more complex than equation (45) could let think, if both decoherence and the relativity of states are taken into account.

Let's start again from equation (44): $|\Psi_{SAE}\rangle = \alpha |+\rangle_z |\uparrow\rangle |E_+\rangle + \beta |-\rangle_z |\downarrow\rangle |E_-\rangle$. This state is the relative (for the observer) state of entanglement between the system, the apparatus and the environment after the interaction between them, written in the preferred basis selected through the Hamiltonian of interaction between the apparatus and the environment. By using the hanging-up mechanism directly applied to equation (45): $|\Psi_{SAEO}\rangle = \alpha |+\rangle_z |\uparrow\rangle |E_+\rangle |☺\rangle + \beta |-\rangle_z |\downarrow\rangle |E_-\rangle |☹\rangle$, and concluding directly that either $\widetilde{☺}$ or $\widetilde{☹}$, we have cut short the decoherence process that is needed to suppress the possibility to observe macroscopic interferences. We have used the hanging-up mechanism exactly like the reduction postulate was used in the measurement theory before decoherence. We could as well have started from $|\Psi_{SA}\rangle = \alpha |+\rangle_z |\uparrow\rangle + \beta |-\rangle_z |\downarrow\rangle$ and get the entangled state with the observer $|\Psi_{SAO}\rangle = \alpha |+\rangle_z |\uparrow\rangle |☺\rangle + \beta |-\rangle_z |\downarrow\rangle |☹\rangle$ to conclude that the consciousness of the observer hangs-up to one of the branch and becomes either $\widetilde{☺}$ or $\widetilde{☹}$. No use of the environment has really been made in this presentation. The fact that we got a correct conclusion is only the reflection of the fact that in general, the reduction postulate can be considered as a shortcut to the decoherence processs, giving correct results when the state vector is written in the preferred basis considered as already known (which is the case for empirical reasons when we use an apparatus built for measuring a specific observable). Now this is not strictly correct and in particular this has a consequence that we want to avoid: it cancels any possibility to observe (even in principle) the interferences coming from the fact that the universal state vector remains superposed. This is precisely what led us to discard the many-worlds interpretation for the benefit of the initial Everett's version.

The correct way to describe the hanging-up mechanism is the following. We start from equation (44): $|\Psi_{SAE}\rangle = \alpha |+\rangle_z |\uparrow\rangle |E_+\rangle + \beta |-\rangle_z |\downarrow\rangle |E_-\rangle$. We assume here that the universe is limited to the system, the apparatus, the environment and the observer. Then equation (44) is the most general state vector that the observer can assign to the universe (herself excepted) after the interaction. Let's notice that this equation could be written in any basis. Then a choice has to be made about what kind of future measurements will not be performed on the universe. This choice can come either from physical limits (compulsory for the observer, such for example the physical impossibility to use an apparatus bigger that the universe) or from a deliberate choice of the observer. Then, a partial trace on the density matrix

$\varrho_{SAE} = |\Psi_{SAE}\rangle\langle\Psi_{SAE}|$ on the degree of liberty that will not be observed must be done to get a reduced density matrix which describes the (still entangled) universe that is accessible to observation. This is only at this step that the decoherence process gives a diagonal density matrix. Now the hanging-up mechanism says that the consciousness of the observer hangs-up to one of the states of the basis in which the reduced density matrix is diagonal with a probability that is given by the diagonal elements. In the case of the measurement of the spin we used, under the assumption that we do not observe the environment, that leads to exactly the same result than the simplified presentation. Now, if we change the assumption about what it is possible to observe in principle (for example assuming that we can measure correlations with the environment), we will trace off different degrees of freedom and will get a new reduced density matrix. It will then become possible that the consciousness of the observer hangs-up to states that show some interferences between the components of the reduced density matrix that we got previously through more stringent restrictions about what can be measured. That means that the branch to which the consciousness hangs-up depends on what is unobservable for the observer and that all the components of the universal entangled state vector are remaining present as it is the case in the initial version of Everett's interpretation.

In the vast majority of cases, this second formulation of the hanging-up mechanism is equivalent to the first one. Nevertheless, the second formulation has two advantages. First, it avoids cutting branches as if once an observation has been made by an observer, the world relative to this observer was no more in a superposed entangled state. Second, it lets decoherence play its full role in providing the observer with the only branches that are accessible for the hanging-up of her consciousness, depending on what is traced off. Hence, the hanging-up mechanism is a weaker postulate than the reduction postulate since it is not in any state of a superposition that the consciousness can be projected but only in those allowed by the decoherence process.

Bohr's claim according to which the value of an observable belongs not to the system but to the whole composed of the system plus the apparatus becomes now easy to understand and even obvious[35]. Through the hanging-up mechanism the value that is measured is not something objective attached to the system but is only the result of the fact that the consciousness of the observer hangs-up to one of the possible branches of the entangled state vector written in the preferred basis which, in turn, can be defined only when the apparatus and its interaction with the environment have been chosen. Therefore the very concept of an objective value attached only to a system independently of any apparatus is meaningless. It is even clearer if we remember how this value is obtained. What decoherence and the hanging-up mechanism say is that the observer's consciousness can only be aware of one of the branches written in the preferred basis. But actually, what the observer's consciousness is aware of is only the macroscopic state of the apparatus because this is what she observes during a measurement (the microscopic system is of course not directly observable). Hence, the value attributed to the observable of the system is only a deduction made by the observer according to the following reasoning: a) I see this macroscopic state of the apparatus, b) this state is correlated to this eigenstate of the microsystem in the branch of the entangled vector state, c) hence, the system is in this eigenstate, d) so the observable has the eigenvalue associated to this eigenstate. But this reasoning is possible only if the observer has gone through all the process of decoherence and hanging-up which is not possible without an apparatus. From the point of view of Convivial Solipsism, Bohr's claim should even be extended to include the observer.

---

[35] Of course, I do not pretend that Bohr had Convivial Solipsism in mind when he claimed that. His claim was a useful assumption helping him to fight against the EPR argument. Convivial Solipsism gives a good reason to understand why Bohr was right even though he probably would not have liked it.

### 8.4. EPR and non locality

As Fine says [51]:

"*Of course it may also be possible to break the EPR argument for the dilemma plausibly by questioning some of its other assumptions (e.g., separability, the reduction postulate, the eigenvalue-eigenstate link, or a common assumption of measurement independence). That might free up the remaining option, to regard the theory as both local and complete. Perhaps a well-developed version of the Everett Interpretation would come to occupy this branch of the interpretive tree*."

Convivial Solipsism could be regarded as this kind of version of Everett's interpretation that Fine has in mind even though the hanging-up mechanism needs to be added. Now, QBism, the relational interpretation and some pragmatist interpretations [71] claim that in their framework the EPR paradox vanishes and that non locality does not exist. But the reasons for that are not exactly the same and vary according to the interpretation. I present here why this is the case in Convivial Solipsism. The EPR paradox comes from the assumption that when Alice and Bob do their measurement, each one on one particle, then the result of their measurement is valid instantaneously for any other observer. In case of an initial singlet state, when Alice finds the value "up" on the first particle, it is assumed that this value is "down" for the second particle immediately for herself and for Bob. The consequence is that it seems that at the very moment when Alice does her measurement, everything is determined both for herself and Bob. Indeed when she asks Bob, later, which value he found on his particle, she will hear "down" and this is true even if Bob's measurement and Alice's measurement are space-like separated. So it seems that the very fact that Alice found "up", determined instantaneously the value "down" for Bob's particle (and vice versa since if the two measurements are space-like separated, no one can be said to be before the other in an absolute way). To be more precise, the reasoning goes like that: when Alice asks Bob which result he got and hears "down", she deduces in retrospect that the value has been "down" as soon as Bob did his measurement. If their measurements are space-like separated this implies a spooky instantaneous action at a distance. This consequence relies on the assumption that state vectors are representing objective physical descriptions of systems and that they are changed by measurements. But if we describe the situation in Convivial Solipsism, this is not true anymore. Indeed, it is only when Alice asks Bob (in the future of her measurement) which value he found that she performs a measurement on Bob and learns what is the result he got on his particle. So, the hanging-up mechanism plays its role and we know that according to it, she will necessarily hear "down" which is the only result in agreement with her own measurement on the first particle. We know also that if she then performs a measurement on Bob's particle, she will find the same "down" value. But that does not mean that the second particle "was already before" in the state "down". Indeed, the hanging-up mechanism is nothing else than the fact that Alice's consciousness hangs-up to the branch corresponding to the value "up" for her particle and "down" for Bob's particle, while the state vector of the systems remains unchanged. These posterior measurements done by Alice are of course time-like separated with Alice's first measurement and moreover, they do not affect physically the state of the system. There is no more "spooky instantaneous action at a distance".[36]

### 9. In summary

Convivial Solipsism (with decoherence) gives satisfying answers to all the questions raised in §6. Of course, the picture it offers can seem a little weird since it means abandoning the idea of an absolute external reality which is the same for everybody. It forbids even considering as meaningful usual sentences that compare the private experience of observers. But abandoning simultaneity was also something shocking for many scientists at the beginning of the twentieth century even though we are all accustomed to it now.

---

[36] Actually, it seems that for QBists, the solution is even simpler. They reject the fact that a probability 1 assignment needs to be backed by an objective fact. Hence, the conclusion of EPR relative to this "element of reality" which can be predicted with probability 1 does not follow anymore. I am not totally convinced because I think that their position about probability 1 raises philosophical problems that should be addressed more carefully after a clarification of the questions mentioned in §3.4.3.

If we let aside the extreme instrumentalist positions which say that the metaphysical questions are not to be considered, it is now possible to understand the appeal of many interpretations that have been previously proposed even if no one succeeded in providing a coherent global answer to all the questions. This appeal comes from the fact that they all share a part of the whole story:

- The Copenhagen interpretation correctly stressed the fundamental role of the experiment and the contextual aspect of the measurement but failed to identify (or at least to make clear) the role of the observer.

- Wigner, London and Bauer correctly noticed that it is impossible to give a coherent interpretation of quantum mechanics without mentioning consciousness. But they wanted it to play a role that is not acceptable.

- Everett in the initial version is close to a coherent picture but, even if it would be unfair to blame him for that, his interpretation needs to be complemented by decoherence. Moreover, the many attacks against his position come from the unclear aspect of this huge multiplication of states of consciousness of the observer and from the difficulty to give any meaning to probabilities and to recover the Born rule.

- The relational interpretation shares with QBism and Convivial Solipsism the concept of relative states. But, as we saw, the way it deals with measurement is unacceptable.

- QBism, conceived as an instrumentalist interpretation has the merit to clearly state when the reduction postulate must be used. But, plunging into more philosophical questions leads to many unanswered questions. Moreover, without any mention to decoherence, QBism is unable to explain the classical appearance of the world that is nevertheless part of each agent's private experience.

Convivial Solipsism gathers Everett's initial position, decoherence and relativity of states in a coherent whole that allows giving answers to the main questions that have been raised at the beginning of this paper. Of course, we must recognize that the image it gives is very unfamiliar and that it is even further from the usual scientific realism than any other proposed interpretation. But, isn't it the case that quantum mechanics has already accustomed us to very strange things?

**Acknowledgement**

I am indebted to Bernard d'Espagnat whom I wish to thank first for many enlightening discussions and for comments he made on a preliminary version of this paper.
I equally want to thank the participants to the Colloquium "Quantum Antinomies and Reality" in June 2015 at the "Fondation des Treilles", specially Michel Bitbol, Caslav Bruckner, Jan Faye, Chris Fuchs, Rom Harre, Richard Healey, Patricia Kauark, Franck Laloë, Jean Petitot, Thomas Ryckman, for useful discussions on this paper, during the colloquium and after.
I also thank David Mermin and Rüdiger Schack for exchanges helping me to clarify my analysis of QBism and Lev Laidman for his comments on my description of the "many worlds" interpretation.